\documentclass[useAMS,usenatbib]{mn2e}

\usepackage{aas_macros}
\usepackage{graphicx}
\usepackage{amssymb}
\usepackage{amsmath}
\usepackage{color}
\newcommand{\Ms} {$\rm{M_{\odot}}~$}

\usepackage{hyperref}
\begin{document}

\title[Shape and Spin of Minihaloes]
{Shape and Spin of Minihaloes: From Large Scales to the Centres}
\author[Druschke et al.]{Maik Druschke$^{1}$\thanks{E-mail: 	
vu412@ix.urz.uni-heidelberg.de},
Anna T. P. Schauer$^{1}$,
Simon C. O. Glover$^{1}$, Ralf S. Klessen$^{1}$\\
$^{1}$ Universit\"at Heidelberg, Zentrum f\"ur Astronomie, Institut f\"ur Theoretische
Astrophysik, Albert-Ueberle-Str. 2, 69120 Heidelberg, Germany\\}

\pagerange{\pageref{firstpage}--\pageref{lastpage}} \pubyear{2018}

\maketitle

\label{firstpage}

\begin{abstract}
The spin and shape of galaxies at the present day have been well-studied both observationally and theoretically. At high redshifts, however, we have to rely on numerical simulations. In this study, we investigate the shape and spin of minihaloes with masses of $M \sim 10^5$--$10^7 \, {\rm M_{\odot}}$ which are of particular interest as they are the sites where the first stars in the Universe form. We analyse a large sample of these minihaloes, selected from a high resolution cosmological simulation. The first minihaloes form at $z \simeq 24$ and by the end of the simulation at $z \simeq 14$ our sample includes $\sim 9000$ minihaloes. 
We find that the spin parameter of the minihaloes follows a log-normal distribution with minimal dependence on redshift. Most minihaloes are prolate, but those formed at the highest redshifts are more prolate than those formed at lower redshifts. On the scale of the virial radius, there is a good correlation between the shape and spin of the gas and that of the dark matter. However, this correlation breaks down in gas which is cooling and undergoing gravitational collapse. We show, contrary to previous assumptions, that although the direction of the spin of the central dense gas correlates well with that of the halo, the magnitude of the spin of the dense gas is uncorrelated with that of the halo. Therefore, measurements of the spin of minihaloes on large scales tell us little about the angular momentum of the gas responsible for forming the first stars.
\end{abstract}

\section{Introduction}
The leading cosmological model that describes the evolution of our Universe from the Big Bang to the present day is the so-called Lambda Cold Dark Matter model ($\Lambda $CDM). In this model, most of the matter content of the Universe is dark (i.e.\ it does not interact electromagnetically) and dissipationless, with a very low initial velocity dispersion. 
It dominates the gravitational potential on large scales (entire galaxies, galaxy clusters etc.), with gas and stars generally dominating on smaller scales. Another important feature of the $\Lambda $CDM is that structure formation in this model is hierarchical: low mass dark matter haloes, or ``minihaloes'' form first, with more massive structures forming later due to ongoing mergers and accretion.

Because the gravitational influence of the dark matter dominates on large scales within galaxies, there is consider interest in understanding how the details of the dark matter distribution affect the distribution of the visible matter. In particular, we would like to know how the shape and spin of the visible matter in a given galaxy relate to the shape and spin of its dark matter halo. 

The shape and spin of galaxies at low redshifts has been studied extensively with both observations 
and simulations \citep[see e.g.][among many others]{Warren, Teklu}. 
However, less work has been done on characterising the properties of the small dark matter haloes in which the first Population~III (Pop~III) stars form. These first objects are minihaloes of masses of $\sim 10^{5-7}$\,\Ms \citep[see e.g.\ the reviews by][and references therein]{volkerreview13,glov13}. They are too distant and too faint to observe directly with current telescopes and we therefore have to rely on simulations to increase our understanding of these objects. 

Several studies have investigated the shape and spin of high-redshift minihaloes using simulations containing only dark matter \citep{Jang,Davis2009,Davis2010,Sasaki}. However, only two studies have investigated the effects on the minihalo properties of including gas \citep{Souza,Hirano}, even though it is known from studies of lower redshift, more massive haloes that gas cooling can lead to clear changes in the halo shape \citep{Kazantzidis}. 

Moreover, these previous studies primarily focussed on the properties of the minihaloes on scales comparable to their virial radii, and did not explore how the shape and spin of the gas distribution on smaller scales correlates with the halo-scale distribution. However, this is of great interest for understanding whether there is any link between the halo-scale properties of the gas and dark matter and the mass distribution of Pop~III stars formed in that halo. 

A quantity of particular interest here is the spin of the gas. High resolution simulations of the gravitational collapse of gas in high-redshift minihaloes have shown that following the formation of an initial, low-mass protostar, further infall tends to build up a large, gravitationally-unstable disk surrounding this protostar \citep[see e.g.][]{stacy10,clark11,get11,sm11,stacy12,Hirano}. Fragmentation of this disk occurs rapidly, and the complex interplay of the fragments  profoundly influences the further evolution of the system.

Simplified models of the evolution of Pop~III accretion disks in the presence of stellar feedback \citep[e.g.][]{McKee} suggest that one of the most important parameters determining the final outcome is the angular momentum of the gas on small scales, as quantified by the ratio of the rotational velocity of the inflow to the Keplerian velocity, $f_{\rm Kep}$, evaluated at the sonic point of the inflow. If this is related in a straightforward fashion to the spin of the halo on large scales (as assumed in the model of \citealt{Souza}), then it may be possible to predict the Pop~III IMF in a large sample of minihaloes without needing to carry out computationally costly simulations of star formation at extremely high resolution in each minihalo. However, it remains to be established whether there is indeed any link between the large-scale spin and small-scale angular momentum in minihaloes.

The goal of this work is to gain a better understanding of the spin and shape parameters in high redshift minihaloes. We compare these parameters evaluated for the halo as a whole with those characterising the dark matter and gas components individually and also investigate how they vary as a function of scale within the cold, collapsing gas that will ultimately form stars.

Our paper is structured as follows. We start with a brief description of our simulation in Section~\ref{Simulation} and discuss our analysis methods in Section \ref{Analysis}. The main results of this paper are presented in Section~\ref{Results}. In detail, Section~\ref{Statistical properties} focuses on the time evolution of the shape and spin parameter over the redshift range from $z=24 $ to $z=14 $. Section \ref{Correlation with gas and dark matter} discusses the individual correlations of these properties between dark matter, gas and the full halo. Furthermore, we investigate in Section \ref{Cold gas} the correlation between cold, dense central gas and the entire halo. Finally, we summarise our results in Section \ref{Conclusion}.
\section{Simulation}\label{Simulation}
The simulation analysed here is run v0 from \citet{anna18}. In the interests of brevity, we give only a few important details here and refer the interested reader to \citet{anna18} for further information. Our simulation is initialised at redshift $z=200$ and runs down to redshift $z=14$, assuming a $\Lambda$CDM cosmology and cosmological parameters derived from \cite{planck15}: $h = 0.6774$, $\Omega_0 = 0.3089$, $\Omega_\mathrm{b} = 0.04864$, $\Omega_\Lambda = 0.6911$, $n = 0.96$ and $\sigma_8 = 0.8159$. The dark matter initial conditions are created with MUSIC \citep{hahn11}, using the transfer functions of \cite{eh98}. The gas is assumed to follow the dark matter distribution initially. 

We use the moving mesh cosmological hydrodynamics code {\sc arepo} \citep{arepo} to perform the simulation. The gas cells are created on an unstructured grid from a Voronoi tessellation and are reconstructed on the hydrodynamical time step \citep{mocz15}. The gas can flow from grid cell to grid cell, and in addition, the cells themselves can move. Time integration is achieved with a Runge-Kutta integration scheme \citep{pakmor16}. Gravity is included for gas cells and dark matter particles using the TreePM method \citep{gadget2} with a hierarchical oct-tree \citep{bh86} for short-range forces, and a Fast Fourier Transform method for long-range forces. 

A primordial chemistry network and cooling function is included into \textsc{arepo} that evolves the species H, H$^{+}$, H$^{-}$, D, D$^{+}$, 
He, He$^{+}$, He$^{++}$, e$^{-}$, HD, H$_{2}$, and H$_{2}^{+}$. 
It is an  updated version of \citet{hartwig15a}, based on previous work 
by \citet{gj07}, \citet{ga08}, \citet{cgkb11} and \citet{glover15}. Further details can be found in \citet{anna17b, anna18}.

The simulation has a high resolution throughout its entire volume. 
We use 1024$^3$ dark matter particles and initially 1024$^3$ gas cells 
for a cosmological box size of $($1\,cMpc/$h)^3$, where c is short for comoving. This results in a dark matter particle mass of $99 \, {\rm M_{\odot}}$ and an average initial gas cell mass of $18.6 \, {\rm M_{\odot}}$. The softening length is set to 20cpc/$h$. The gas cells are allowed to refine and de-refine as necessary to keep their masses close to the initial cell mass. For high density gas, we use a Jeans refinement criterion. In order to keep the time step in the simulation reasonable, refinement is stopped if the cell volume decreases below $V_{\rm min} = 0.1 \: h^{-3} \, {\rm cpc^{3}}$. 

The minihaloes analysed in our study are found via a friends-of-friends algorithm. All dark matter particles that are separated by a distance closer than the so-called linking length are associated with the halo \citep{fof}. We use the standard linking length of $b = 0.2$ in units of 
the mean interparticle separation, which yields haloes with a typical overdensity of $\sim$200, similar to the value used for defining the virial radius.

We need to ensure that our haloes are well resolved. Many studies use a threshold of at least 100 particles for their selection of minihaloes \citep[see e.g.][]{wise12}. However, \citet{Sasaki} show that this is not enough to allow us to measure the shapes of haloes accurately. We therefore use a more demanding criterion of a minimum halo mass of $M_\mathrm{min} = 6 \times 10^4 \: {\rm M_{\odot}}$. This corresponds to 500 dark matter particles and a comparable number of gas cells per halo.

\section{Analysis}\label{Analysis}
\subsection{Angular momentum}
We calculate the angular momentum $\vec{J}(R){}$ as a function of radius 
$R \leq R_{\rm vir}$ for each minihalo in our simulation using the expression:
\begin{equation}
\begin{split}
 \vec{J}(R) = \sum\limits_{r_{i}<R} m_{i} \vec{r}_{i} \times \vec{v}_{i}.
\end{split}
\end{equation}
Here, $\vec{r}_i$ is the position of the dark matter particle or the centre of the gas cell relative to the most bound cell within the virial radius, which we take to represent the centre of the halo, $r_{i} = |\vec{r}_{i}|$, $m_i$ is the mass of the particle or cell, and $\vec{v}_i$ is its velocity relative to the centre of the halo. Note that unless otherwise stated, our calculation of the angular momentum and the spin parameter (see below) includes both the gas and the dark matter.
\subsubsection{Definition of the spin parameter}
The angular momentum of our minihaloes varies significantly with halo mass, since more massive haloes contain more gas cells and dark matter particles, and typically have larger rotational velocities. Therefore, when dealing with haloes spanning a wide range of masses, it is more convenient to work in terms of a dimensionless spin parameter. The standard definition of this parameter, first introduced by \cite{Peebles}, is simply
\begin{equation}
\lambda = \dfrac{\vert J \vert \vert E \vert^{\frac{1}{2}}}{GM^{\frac{5}{2}}},
\end{equation}
where $E$ and $M$ are the potential energy and mass of the halo, $J$ is its angular momentum, and $G$ is the gravitational constant.

However, as pointed out by \cite{Bullock}, the total energy $E$ can be difficult to calculate accurately when dealing with haloes in crowded regions, or with subsets of particles within haloes. Therefore, in our analysis, we use a modified definition of the spin parameter given by 
\citep{Bullock}
\begin{equation}\label{LambdaGasDM}
\lambda^{\prime}_{i}(R)= \dfrac{\vert J \vert_{i}(R)}{\sqrt{2} R\, M_{i}(R)V_{\mathrm{circ}}(R)}. 
\end{equation}
Here, $i$ denotes the component of the halo that we are interested in (gas, dark matter, or the total matter content), $J_{i}$ and $M_{i}$ are the total angular momentum and mass of that component within a sphere of radius $R$, and $V_{\mathrm{circ}}$ is the circular velocity of the halo
\begin{equation}
V_{\mathrm{circ}}(R) = \sqrt{\dfrac{G M(R)}{R}},
\end{equation}
where $M$ is the total mass of the halo within a sphere of radius $R$ around its centre [i.e.\ $M(R) = M_{\rm gas}(R) + M_{\rm dm}(R)$]. This modified version of the spin parameter reduces to the original definition in the special case where we have a singular isothermal sphere and $R = R_{\rm vir}$, and even in the more general case, $\lambda^{\prime}$ and $\lambda$ do not differ by more than a few percent when measured at the virial radius.

In our analysis later in the paper, we typically calculate $\lambda^{\prime}$ at the virial radius of the halo, using the notation $ \lambda^{\prime} \equiv \lambda^{\prime}(R_\mathrm{vir})$. However, in Section~\ref{Cold gas}, we are interested in the value of $\lambda^{\prime}$ in gas above some specified number density threshold $n_{\rm thres}$. In this case, we identify the radius of the sphere around the halo centre that contains all of the gas cells with $n \geq n_{\rm thres}$ and then compute the value of $\lambda^{\prime}$ at this radius. 
For example, $\lambda^{\prime}_{\mathrm{100}}$ is the spin parameter of all gas contained in a sphere with radius $r_{\mathrm{100}}$ that reaches out to the farthest gas cell with $n \geq 100{}$\,cm$^{-3}$. 
\subsubsection{Distribution of the spin parameter}\label{Distribution}
The next step after calculating the angular momentum and the spin parameter of a single halo is to investigate the statistical distribution over our minihalo sample. To ensure that we are dealing with well-resolved haloes, we restrict our attention to haloes more massive than $M_{\mathrm{min}} \geq 6.5 \times 10^{4} \, \mathrm{M}_{\mathrm{\odot}}$. These haloes typically contain 500 or more dark matter particles and a comparable number of gas cells.

\begin{figure} 
\includegraphics[width=0.99\columnwidth]{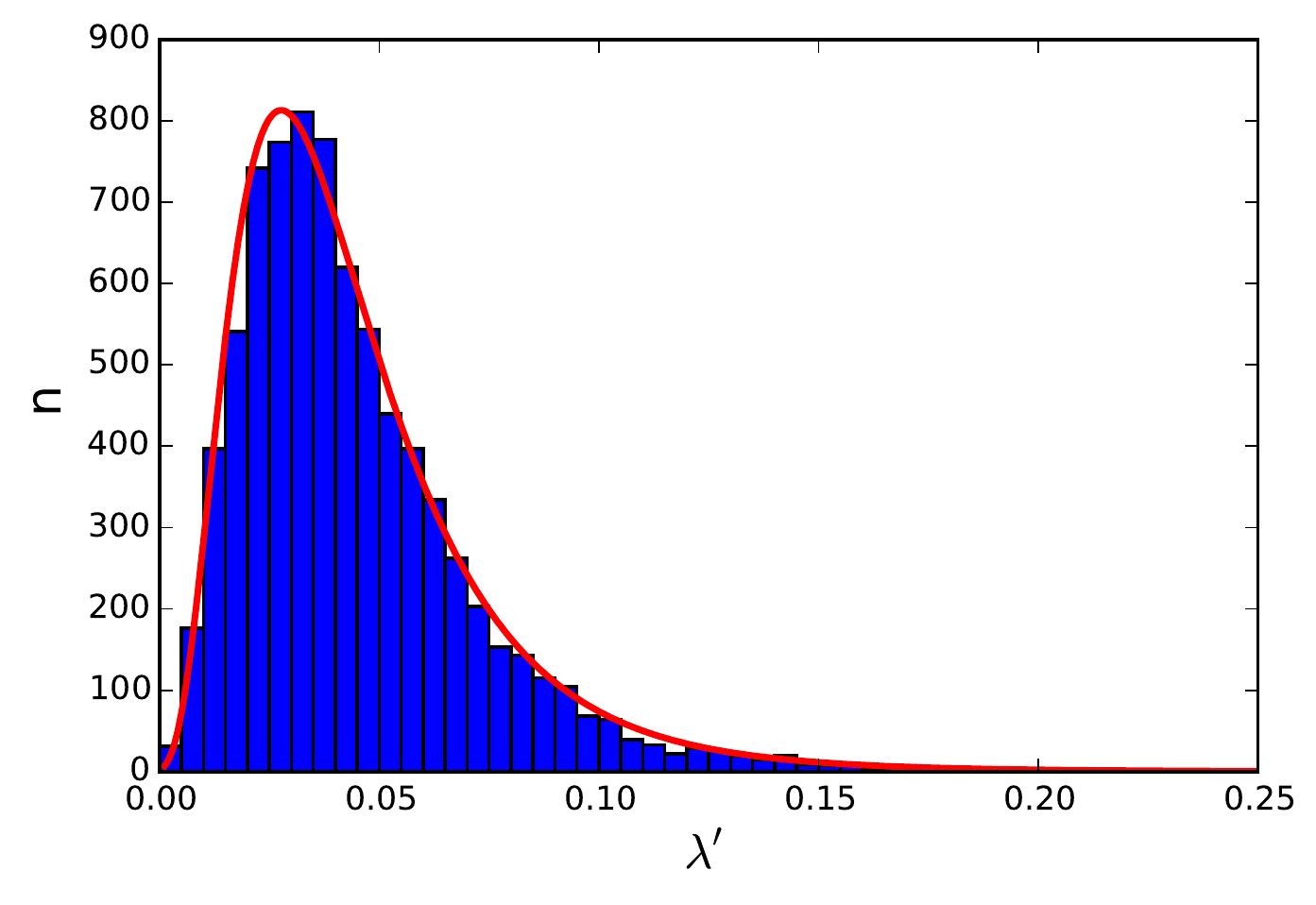}
\caption{Distribution of the spin parameters for all haloes with $M \geq M_{\rm min}$ at a redshift of $z=14$. The red curve is a log-normal fit to the data, with parameters as given in the text.}
\label{fig:LambdadistributionGesamt}
\end{figure}

Our simulation  
contains 9020 haloes with $M \geq M_{\mathrm{min}}$ at our final output 
at $z=14$. In Figure \ref{fig:LambdadistributionGesamt}, we show the distribution of the spin parameters for these haloes.

As one can see by comparing the red line to the blue histogram, the distribution can be well described by a log-normal function, as introduced by e.g.\ \cite{Mo} or \cite{Bullock}:
\begin{equation}
P(\lambda^{\prime})  = \dfrac{1}{\lambda^{\prime} \sqrt{2\pi}\sigma_{0}} \mathrm{exp\left(-\dfrac{ln^{2}\left(\frac{\lambda^{\prime}}{\lambda_{0}}\right)}{2\sigma_{0}^{2}}\right)} .
\end{equation}\label{Lambda}Here, $\lambda_{0}$ and $\sigma_{0}$ are the location parameter and shape parameter of the distribution. 

In our simulation at $z=14$, this log-normal fit peaks at $\lambda^{'} = 0.0253$. We have estimated the error in this peak value using bootstrap sampling, finding a one sigma value of around $2 \times 10^{-4}$, or a fractional error of around 1\%. At high redshifts, where our sample of minihaloes is much smaller, the error is considerably larger. 

Our recovered peak value is consistent with that found by \cite{Sasaki} at a similar redshift, but is somewhat larger than the value of $ \lambda^{\prime} $ = 0.0184 found by \cite{Souza}, and somewhat smaller than the value of $\lambda = 0.04$ found by \cite{Davis2010}.
\begin{figure} 
\includegraphics[width=0.99\columnwidth]{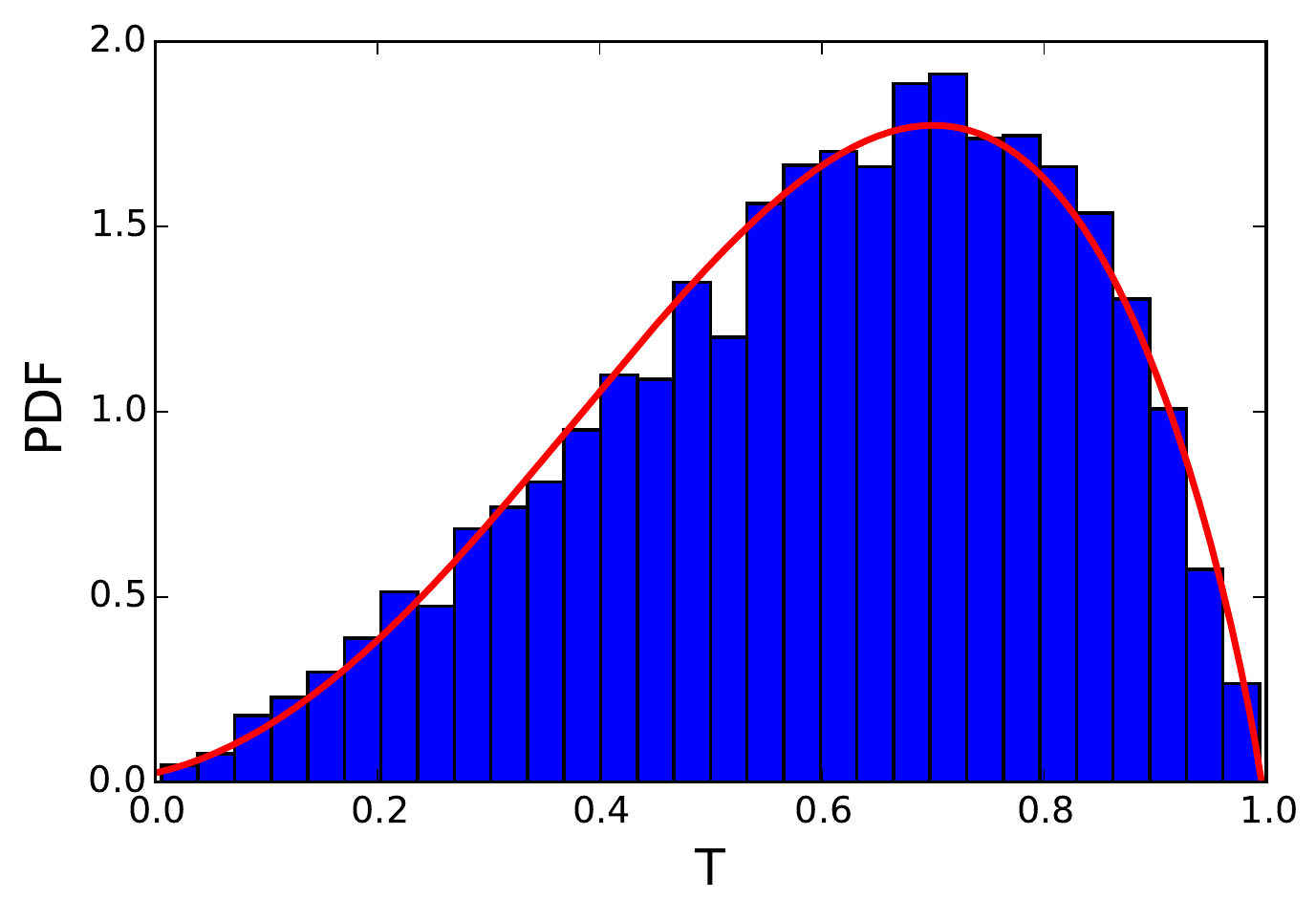}
\caption{Distribution of the triaxiality for all haloes with $M \geq M_{\rm min}$ at a redshift of $z=14$. The data are well described by a beta distribution, indicated by the red curve.}
\label{fig:TriaxdistributionGesamt}
\end{figure}

\subsection{Triaxiality}\label{Triaxiality and sphericity}
In addition to the spin, we also examine the shape of each minihalo, as quantified by its triaxiality.\footnote{We also examined the sphericity of the minihaloes but found that it provides little additional useful information.}
To calculate this quantity, we use all particles and gas cells that belong to the halo inside the virial radius $R \le R{_\mathrm{vir}}$. 
We begin by introducing the inertia tensor $I$, following the definition from \cite{Souza} and \cite{Springel2004}: 
\begin{equation}
I_{jk} = \sum_{i=1}^N m_{i}(r_{i}^{2}\delta_{jk}-r_{ij}r_{ik}) .
\end{equation}  
Here, m$_{i}{}$ is the mass and r$_{i} $  is the distance to the halo centre of the i-th particle or gas cell, and $ \delta_{jk} $ the Kronecker delta. 
The eigenvalues ($ I_{1} $ $ \geq $ $ I_{2} $ $ \geq $ $ I_{3} $) of this inertia tensor then yield the lengths of the three principal semi-axes of the minihalo, $a$, $b$ and $c$, via
\begin{eqnarray}
a &=& \sqrt{\dfrac{5(I_{1}+I_{2}-I_{3})}{2 M}}, \\
b &=& \sqrt{\dfrac{5(I_{1}-I_{2}+I_{3})}{2 M}}, \\
c &=& \sqrt{\dfrac{5(-I_{1}+I_{2}+I_{3})}{2 M}},
\end{eqnarray}
where $M$ is the total mass of the halo inside the virial radius. The axes $a$, $b$ and $c$ define an ellipsoid  with $a \geq b \geq c$. 
\begin{figure*}
\includegraphics[width=1.95\columnwidth]{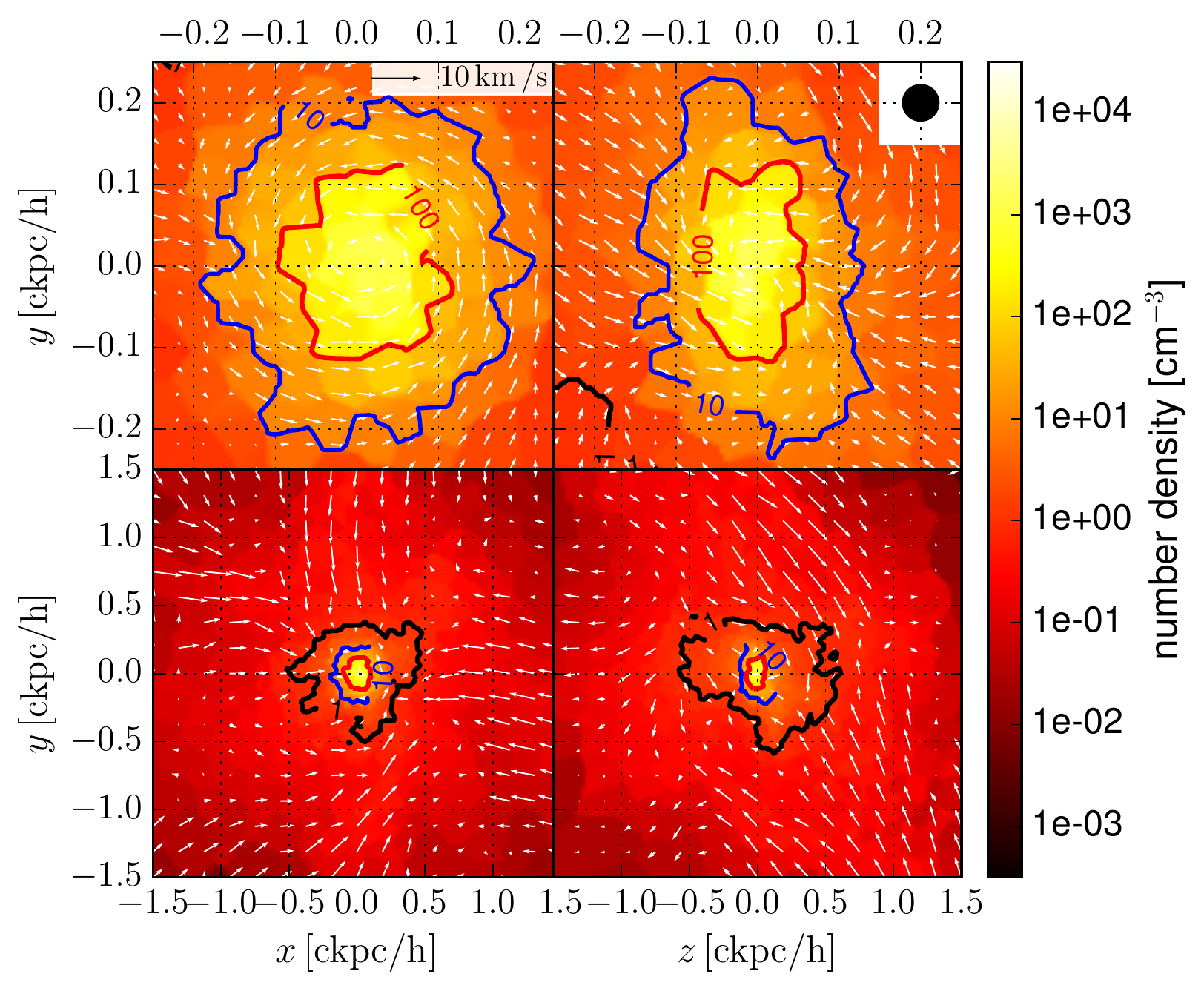}
\caption{Example of a typical minihalo with $\lambda^{\prime} = 0.0249$ and $M = 	1.93\times 10^{6}\, \mathrm{M}_\odot$ at $z = 14$. The left panels show slices through the minihalo in the $x-y$ plane, while the right panels show slices in the $z-y$ plane. The slices are colour-coded by number density, and contours indicating number densities of 1, 10 and 100$\ \mathrm{cm}^{-3}$ are also shown. In the upper panels, the region within a box of side length $0.5$ ckpc/$h$ is shown, while in the lower panels a larger region of side length $3.0$ ckpc/$h$ is shown. The arrows indicate the direction and magnitude of the gas velocity.
For reference, we show a black arrow corresponding to a velocity of $10 \, {\rm km \, s^{-1}}$ in the top left panel. The dark circle in the upper right corner has a radius equal to the gravitational softening length.} 
\label{fig:Densityplot}
\end{figure*}
The triaxiality is defined by \cite{Franx} as
\begin{equation}
T = \dfrac{a^{2}-b^{2}}{a^{2}-c^{2}} .  \label{triax-eq}
\end{equation}
If $a = b > c$, then $T = 0$ and the halo is shaped like an oblate spheroid. On the other hand, if $a > b = c$, then $T = 1$ and the halo is shaped like a prolate spheroid. $T$ is therefore a measure of the oblateness or prolateness of the halo.
\subsubsection{Distribution of triaxiality}
%
\begin{figure*}
\includegraphics[width=1.95\columnwidth]{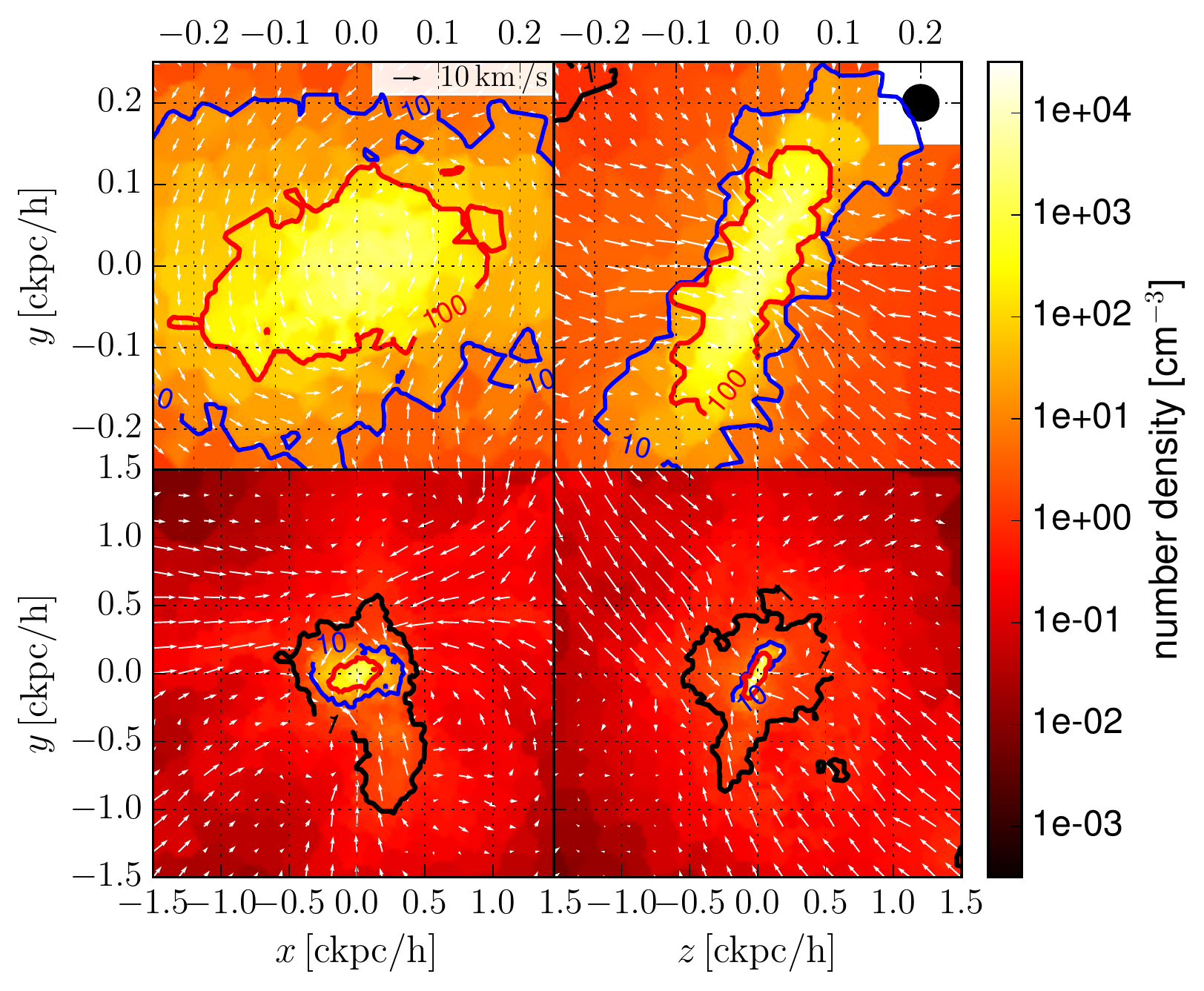}
\caption{As Figure~\ref{fig:Densityplot}, but for a low spin halo with 
$\lambda^{\prime} = 0.0055$ and $M = 	3.03\times 10^{6}\, \mathrm{M}_\odot$ at $z = 14$.} 
\label{fig:Densityplot_lowspin}
\end{figure*}

Figure \ref{fig:TriaxdistributionGesamt} shows the distribution of the triaxialities of all minihaloes with $M \geq M_{\rm min}$ at $z=14$. Although there is a broad distribution of values, we find that minihaloes tend to be more prolate than oblate. 

The histogram can be well described by a beta function:
\begin{equation}
P(T)=\dfrac{\Gamma(a+b)T^{a-1}(1-T)^{b-1}}{\Gamma(a)\Gamma(b)}
\end{equation}
with shape parameters $a$ and $b$, where $T$ is the triaxiality as defined in Equation~\ref{triax-eq}. The gamma function $\Gamma$ ensures normalisation:  
\begin{equation}
\Gamma (n)=\int _{0}^{\infty} t^{n-1} e^{-t}\, \mathrm{d}t .
\end{equation}
The best-fitting beta distribution has $a = 3.35$ and $b = 1.89$ and yields a most probable value for the triaxiality of $T = 0.700$. 

The fact that most high-redshift minihaloes appear to be prolate rather than oblate is in good qualitative agreement with the earlier studies of \citet{Jang} and \citet{Sasaki}, although these dark-matter-only simulations find that the distribution of triaxialities peaks at an even larger value of $T$ than we find here. Studies of the shapes of considerably more massive haloes also find that prolate haloes tend to be more common than oblate haloes \citep[see e.g.][]{Warren,Allgood}.

\section{Results}\label{Results}
We start our analysis by visually inspecting different halo types. 
Figures~\ref{fig:Densityplot} and \ref{fig:Densityplot_lowspin} show slices of number density in the $x$-$y$ and $z$-$y$ planes for a typical halo with $\lambda^{\prime} = 0.0249$ and $M = 1.93\times 10^{6}\, \mathrm{M}_\odot$ and a low spin parameter halo with $\lambda^{\prime} = 0.0055$ and $M = 3.03\times 10^{6}\, \mathrm{M}_\odot$, respectively. In both cases, the state of the gas is shown at redshift $z = 14$. The top row of panels in the Figures shows the gas distribution in the $x$-$y$ and $z$-$y$ planes in the central 500 cpc/$h$ around the most bound cell. The lower row of panels show a similar view for a larger region of size 3000 cpc/$h$ around the most bound cell. Note that in both cases, the coordinate system is chosen so that the net angular momentum of the gas within the virial radius points along the $z$-axis.
The coloured contour levels show number densities of 1 (black), 10 (blue), and 100 (red) in units of $\mathrm{cm}^{-3}$. The black circle in the upper right corner has a radius equal to the gravitational softening length. It is clear that this is considerably smaller than the size of the region with $n \geq 100 \, {\rm cm^{-3}}$, demonstrating that the properties of gas at or below this density are not significantly affected by the gravitational softening. The white arrows represent the velocities of the cells. 

In both cases, it is clear that the dense gas ($n \geq 100\ \mathrm{cm}^{-3}$) has formed a flattened, rapidly rotating structure whose major axis is approximately perpendicular to the axis of rotation. Although these structures are not completely rotationally supported and are too thick to be well described as classical thin disks, it is nevertheless convenient to refer to them as disks, and we will therefore do so in the remainder of this section. We examine the spins of these disks in more detail in Section~\ref{Cold gas} below.

\subsection{Statistical properties}\label{Statistical properties}
\subsubsection{Evolution with time}\label{Evolution with time}
\begin{figure}
\includegraphics[width=0.99\columnwidth]{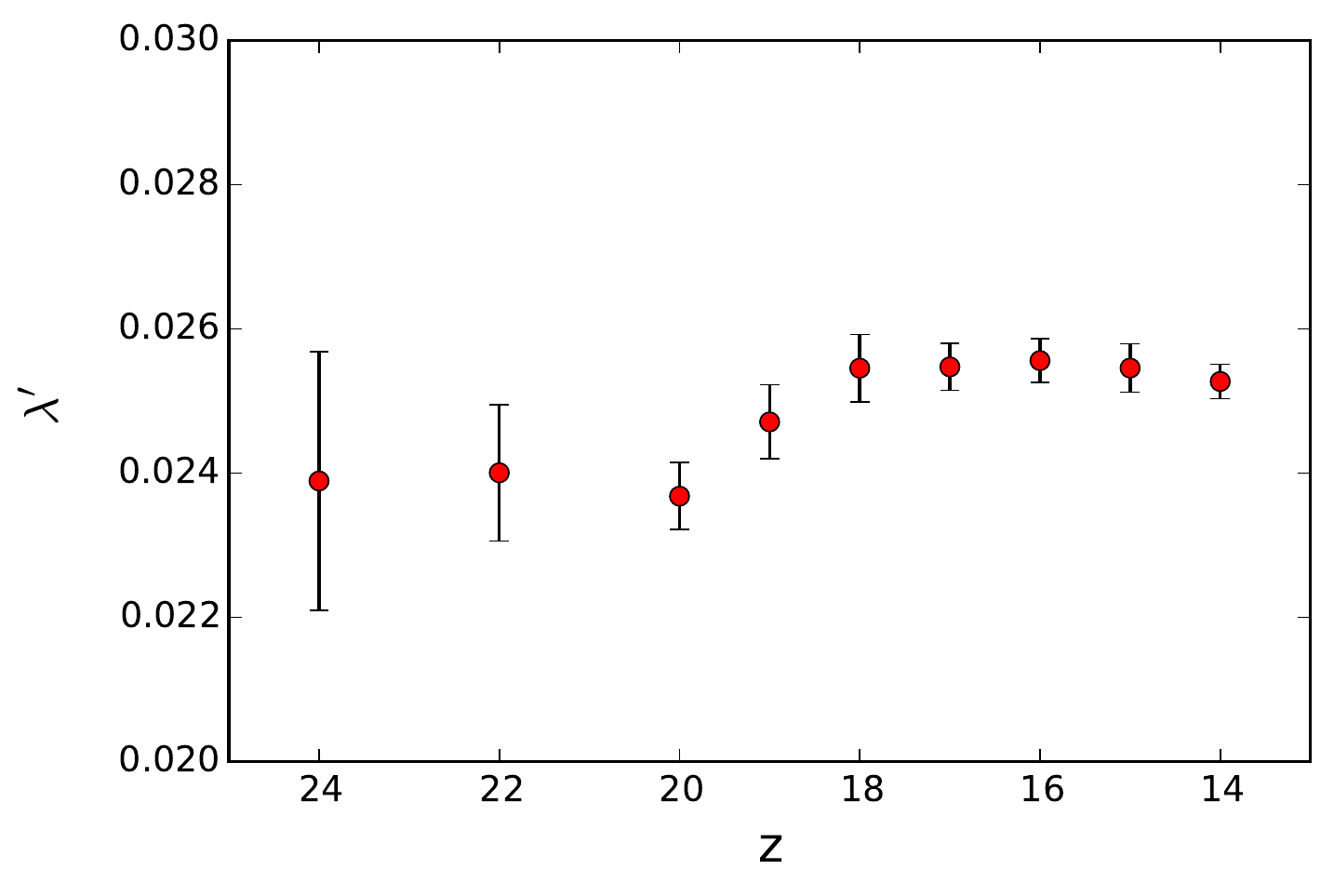}
\caption{Evolution of the $\lambda^{\prime} $ peak values against redshift. The error bars are estimated using the bootstrap method, as described in the text.}
\label{fig:Lambda vs redshift plot}
\end{figure}

First, we want to determine how the minihalo spin parameter distribution evolves with time. 
\begin{figure}
\includegraphics[width=0.99\columnwidth]{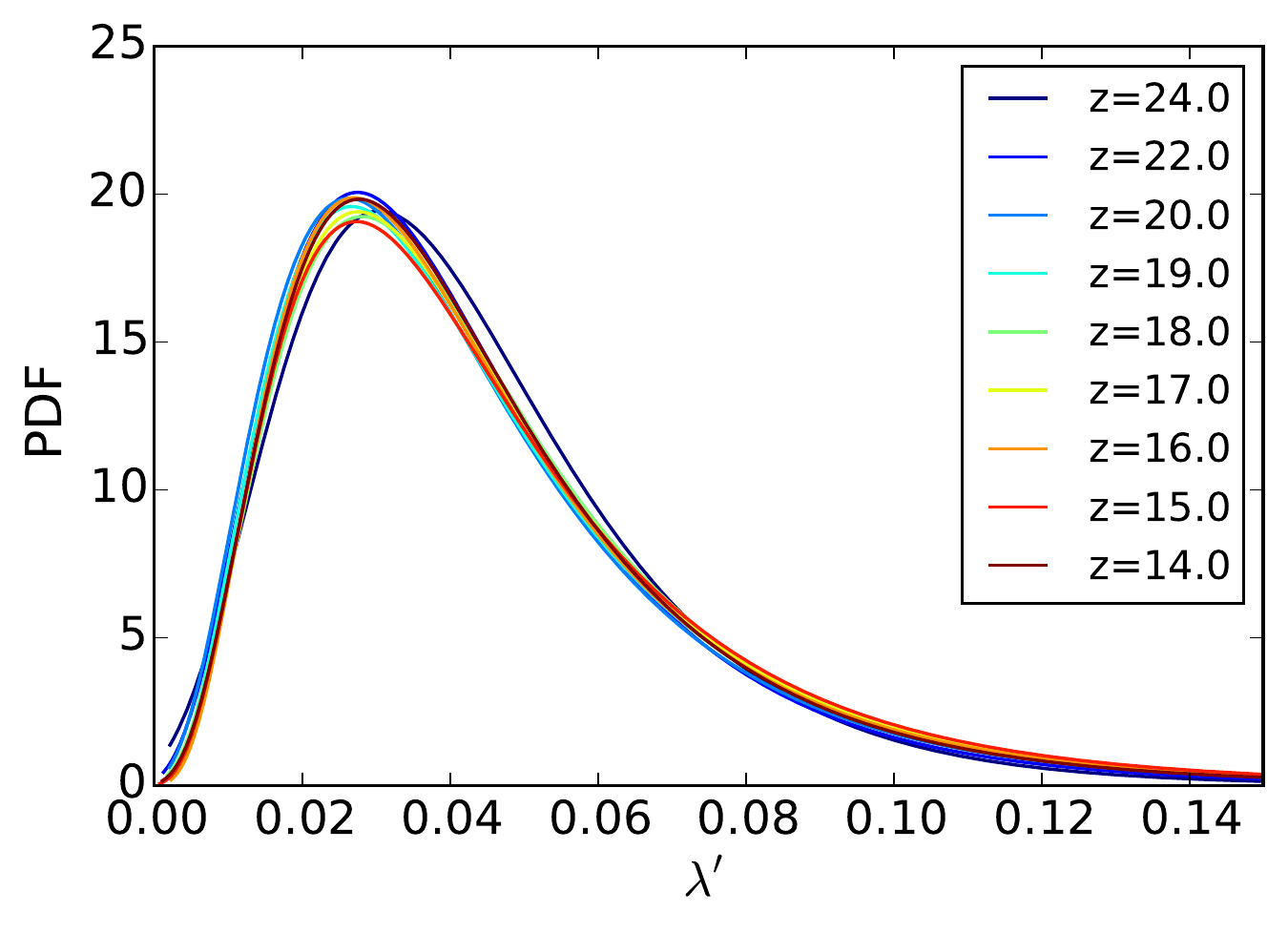}
\caption{Evolution of the $\lambda^{\prime}$ distribution for several redshifts. The colour table increases its blue tone for increasing redshift $z$.}  \label{fig:Lambdadistrall vs redshift plot}
\end{figure}
We have analysed the spin parameter distribution for well-resolved minihaloes (i.e.\ ones with $M \ge M_{\rm min}$) at a number of different redshifts between $z = 14$ and $z = 24$. At all of the redshifts that we have analysed, the distribution is well-fit by a log-normal function. Therefore, in Figure \ref{fig:Lambdadistrall vs redshift plot}, we show for simplicity only these log-normal fits rather than the full distributions.
The peak value of the spin parameter distribution of all haloes as a function of redshift is shown in Figure \ref{fig:Lambda vs redshift plot}. 
There is some hint of an increase in $\lambda^{\prime}$ with decreasing redshift at $z > 18$, although this is barely statistically significant. 
At redshifts $z \le 18$, we find no significant redshift dependence of $\lambda^{\prime}$ on $z$. These results are in good agreement with earlier studies that also found that the spin parameter of high redshift minihaloes depends only very weakly on $z$ \citep{Davis2009,Souza}. 

\begin{figure}
\includegraphics[width=0.99\columnwidth]{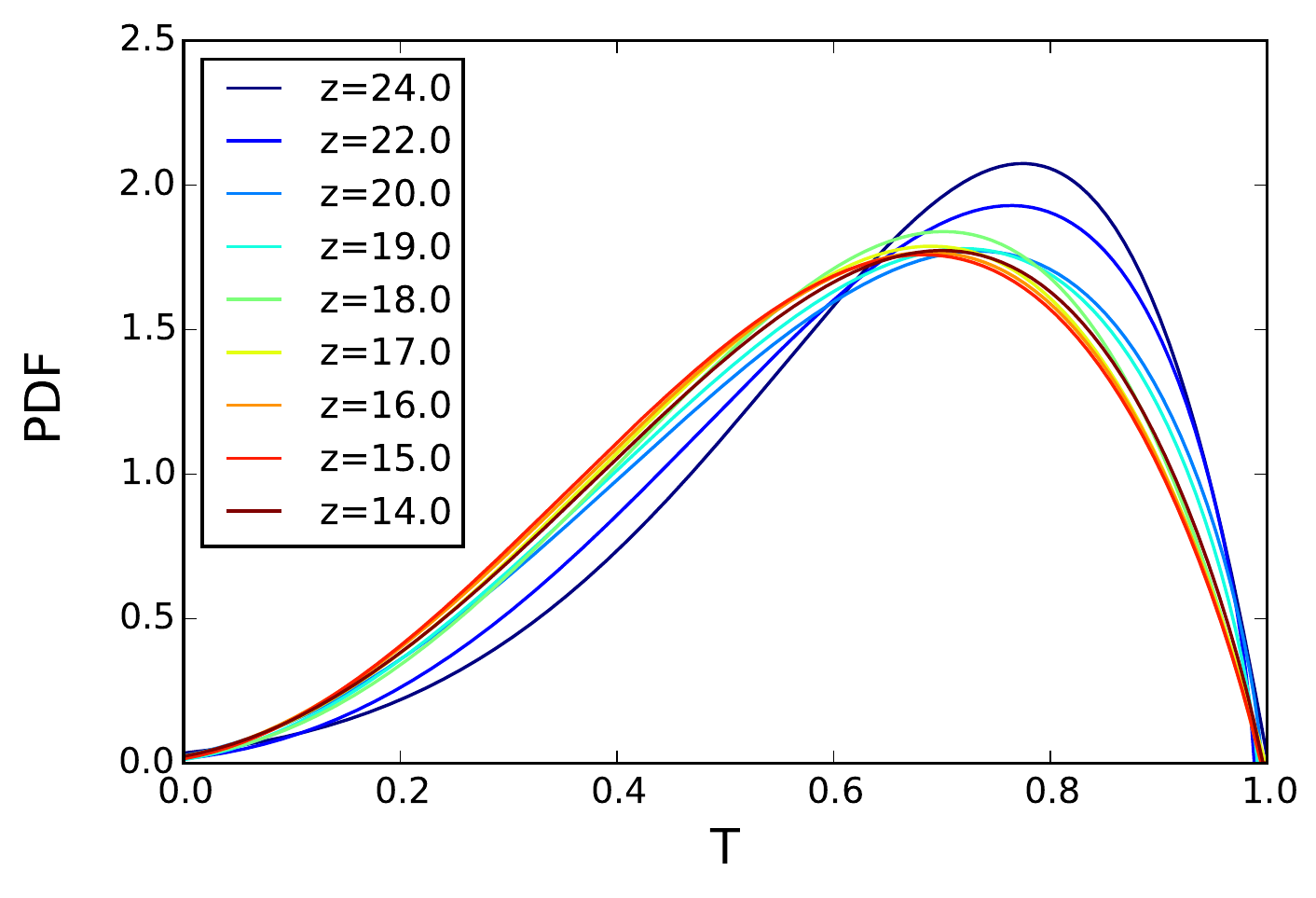}
\caption{Evolution of the triaxiality distribution for several redshifts. The colour table increases its blue tone for increasing redshift z. }  \label{fig:Traixdistrall}
\end{figure}

\begin{figure}
\includegraphics[width=0.99\columnwidth]{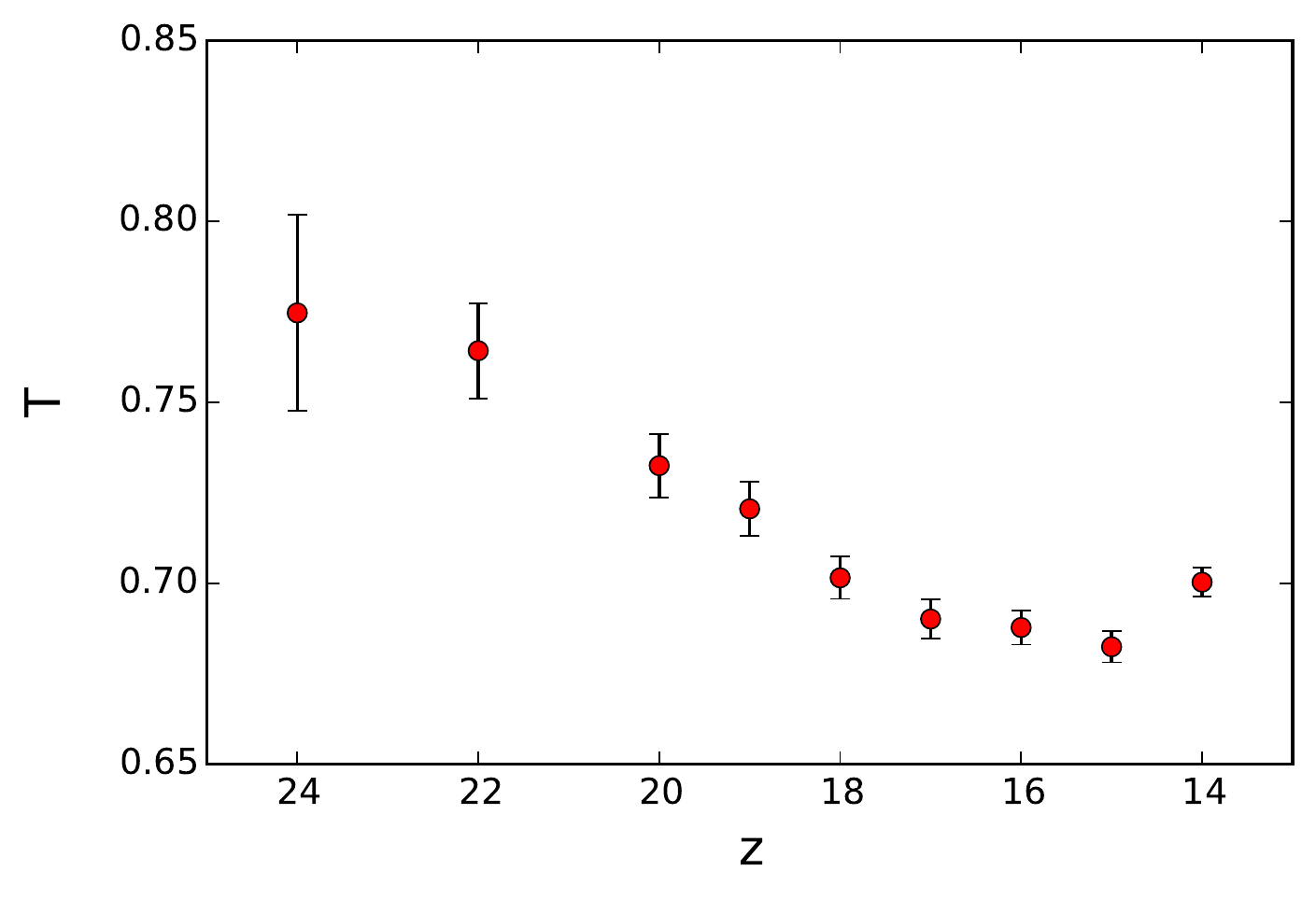}
\caption{Evolution of the $T_{\mathrm{tot}}$ peak values with redshift. The error bars are estimated using the bootstrap method.}
\label{fig:Triaxdistrall vs redshift plot}
\end{figure}

We have also repeated this analysis for the triaxiality. In Figure \ref{fig:Traixdistrall}, 
we show the normalized beta distribution fits for the redshift range from $z=24 $ to $z=14$. 
It is apparent from the Figure that the haloes become less prolate with decreasing redshift. 
This can be seen more directly if we look at the variation of the peak value of the fit as a 
function of redshift (Figure~\ref{fig:Triaxdistrall vs redshift plot}). 
This behaviour is likely a result of the rapid growth of structure at the earliest times: haloes that have recently undergone mergers will tend to be prolate, and the haloes in our high redshift snapshots will have undergone mergers more recently, on average, than those in the lower redshift snapshots. It is notable that similar results hinting at a decreasing triaxiality for decreasing redshift have also been found for much more massive haloes at low $z$ \citep[see e.g.][]{Knebe}.
\subsubsection{Alignment of haloes}
\begin{figure} 
\includegraphics[width=0.99\columnwidth]{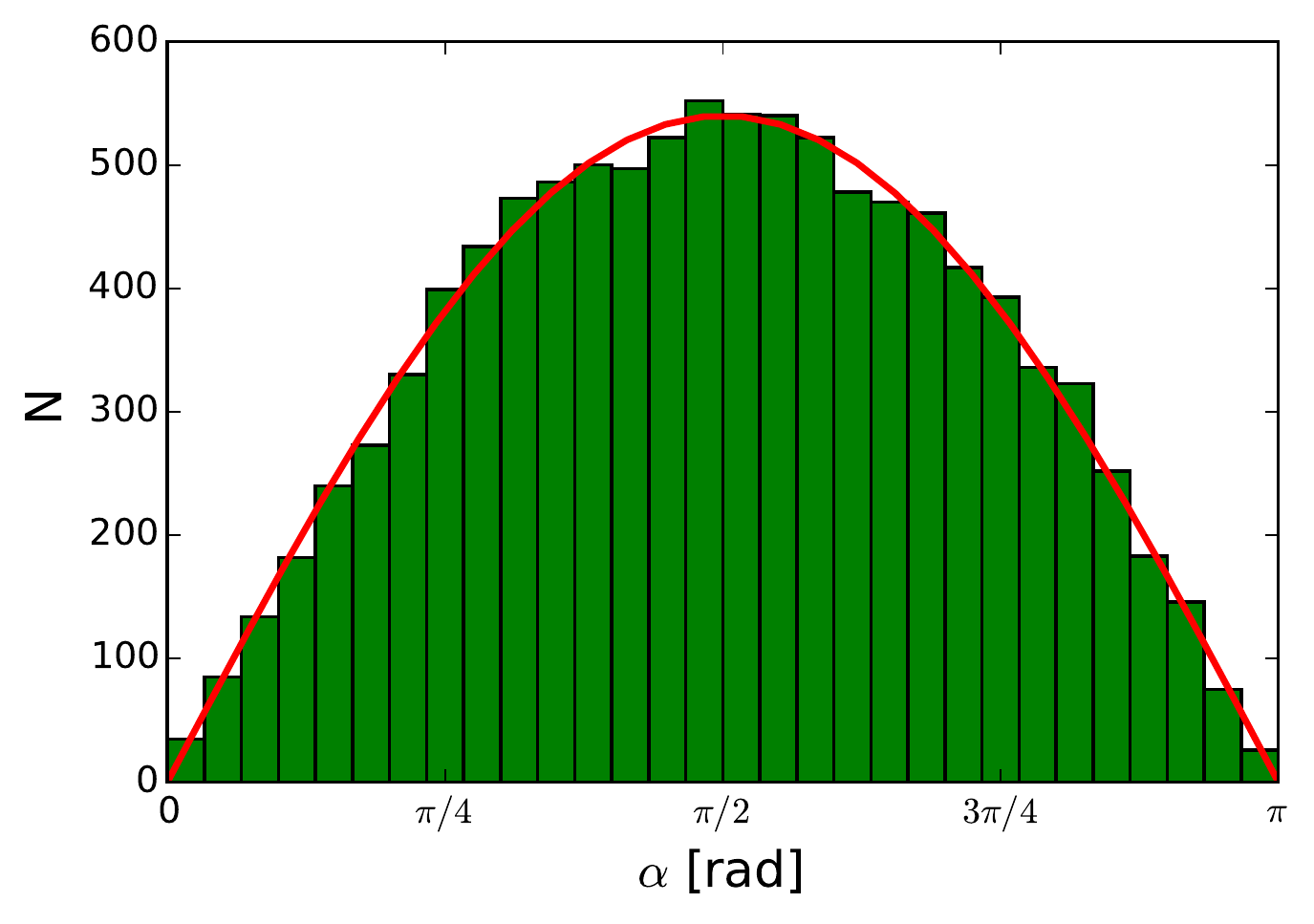}
\caption{Histogram of the angle between the angular momenta of all halo pairs with distance $\le 50$\,ckpc/$h$. The red line shows what one expects for a completely random distribution.}  
\label{fig:HistoalphadistributionGesamt}
\end{figure}

We are also interested in whether the angular momentum of neighbouring haloes is aligned in our simulation. If this was the case, it would hint at an underlying large-scale velocity field. 
Moreover, we expect the vast majority of the minihaloes in our sample to eventually merge into larger galaxies, and it is reasonable to expect that the outcome of these mergers will depend to some extent on the degree to which the spins of the minihaloes are aligned. We therefore study the angle $\alpha{}$ between the angular momentum of halo pairs
\begin{equation}
\alpha_{ij}=\arccos\left(\dfrac{\vec{J_{i}} \cdot \vec{J_{j}}}{\left| \vec{J_{i}}\right|  \left| \vec{J_{j}} \right|}\right).
\end{equation}
As computing $\alpha_{ij}$ is computationally costly when the number of halo pairs is large, we restrict our study to only the first $\sim 1000$ 
most massive haloes at $\mathrm{z}=14$ and compute $\alpha_{ij}$ for all halo pairs with a separation of 50 ckpc/$h$ or less. This yields a total of $\sim 10000$ halo pairs. The histogram of the corresponding angles is shown in Figure \ref{fig:HistoalphadistributionGesamt}, along with the expectations for a random distribution (red line). It is clear that the distribution of angles in our simulation is consistent with a random distribution, and that there is no evidence for any correlation between the spin directions of neighbouring minihaloes on these scales.

\begin{figure} 
\includegraphics[width=0.99\columnwidth]{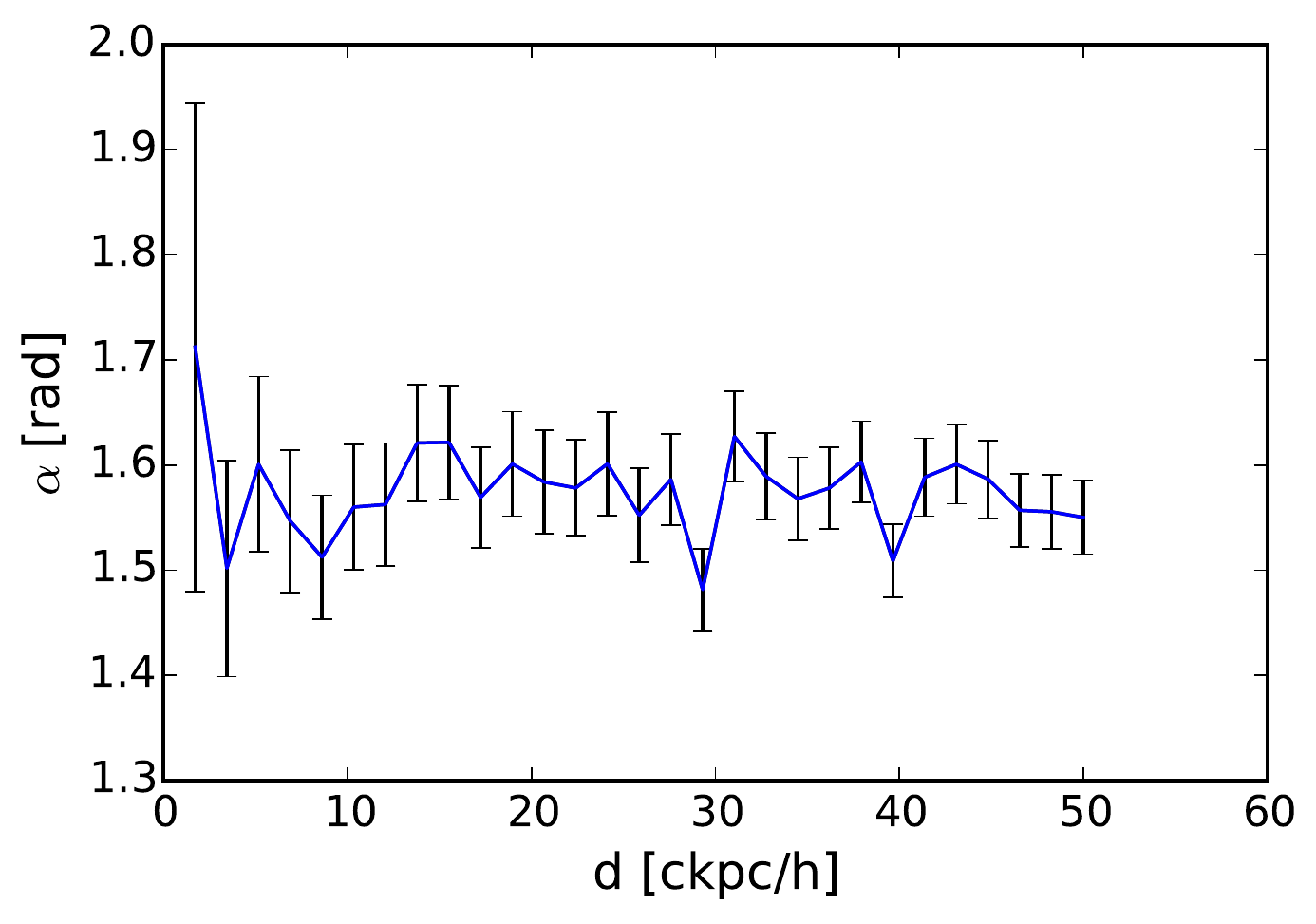}
\caption{The mean angle between the angular momenta of halo pairs with separation $\leq d$. The error bars show the standard deviation. For a completely random distribution, we expect to recover a value of $\pi/2$.}
\label{fig:AngleDistance}
\end{figure}

Our sample of halo pairs is dominated by pairs with our largest allowed separation, 50 ckpc/$h$, and so we have also investigated whether decreasing this maximum separation significantly affects this result. We show in Figure \ref{fig:AngleDistance} the mean angle between the angular momenta of all halo pairs with a separation smaller than $d$, plotted as a function of $d$. The error bars show the standard deviations of the alignment angle. We see that at all but the smallest separations, where there are too few minihaloes pairs to allow us to derive statistically meaningful results, the mean values are consistent with the value of $\pi/2$ expected for a random distribution. 

\subsection{Correlation between the gas and the dark matter}
\label{Correlation with gas and dark matter}
\begin{figure*}
\includegraphics[width=1.99\columnwidth]{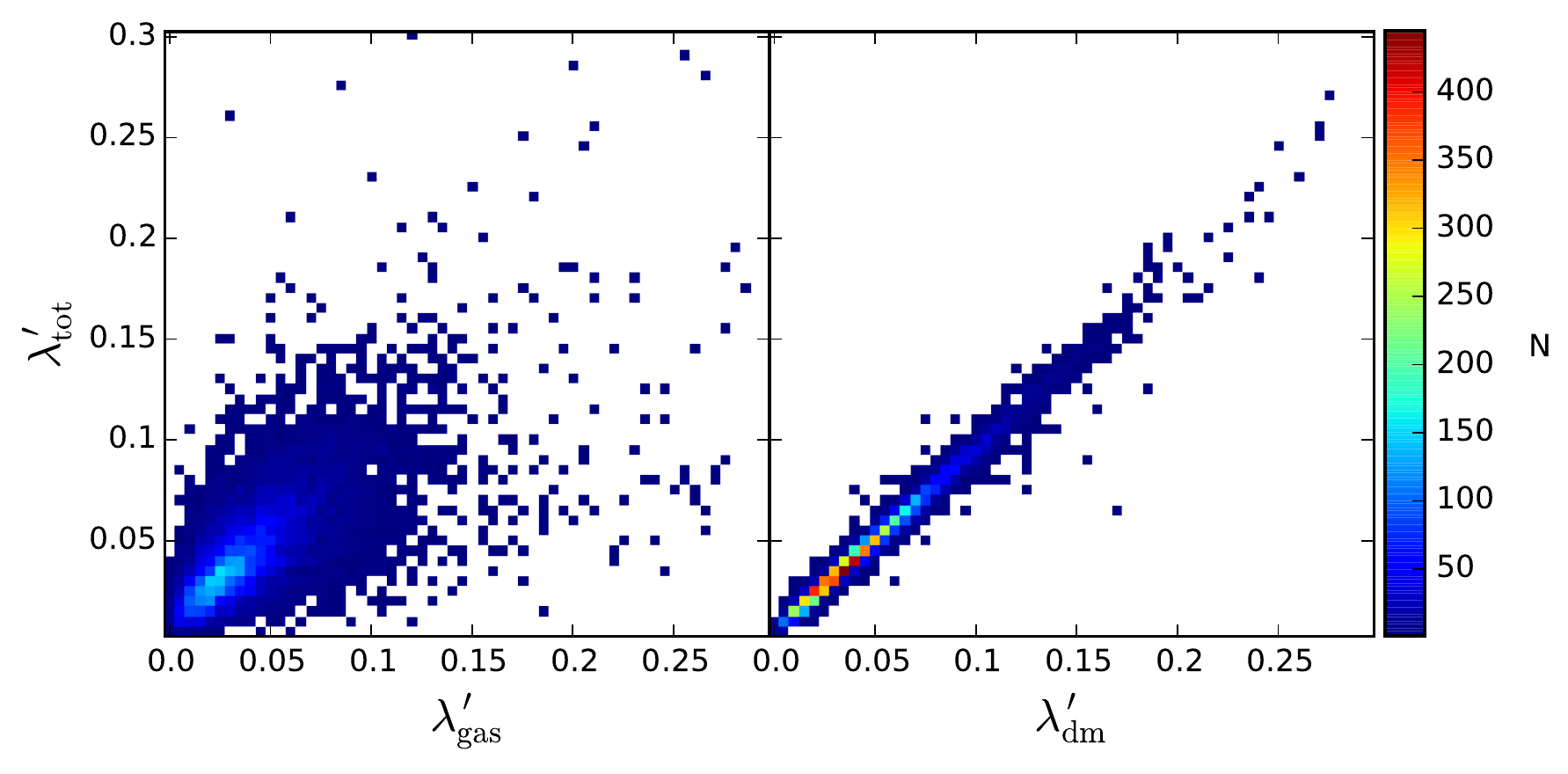}
\hspace{-0.1in}
\caption{Two dimensional histograms for the spin parameters of the entire halo and the dark matter (right panel) and gas (left panel) components. The colour shows the number of minihaloes represented by each pixel. The pixel size is 0.005 $\times $ 0.005.}   
\label{fig:2dhisto_lambda}
\end{figure*}
So far, we have focussed on investigating the spin parameter and shape of all of the matter within each minihalo, drawing no distinction between the gas and the dark matter. However, it is also interesting to examine how well the spin and shape of all of these components considered separately reflects the spin and shape of the halo as a whole. 

In Figure \ref{fig:2dhisto_lambda} we show 2D histograms of the distribution of the spin parameter for gas versus total matter (left panel) and for dark matter versus total matter (right panel). We see that the spin of the entire halo is well correlated with the spin of the dark matter component. This is expected, since the dark matter dominates the total mass of each halo. There is also a clear correlation between the 
spin of the gas and the total spin, but the scatter is much larger, and there is a hint that the gas tends to have slightly larger spin than the dark matter \citep[see e.g.][who find a similar result for lower redshift galaxies.]{Teklu}

\begin{figure}
\includegraphics[width=0.99\columnwidth]{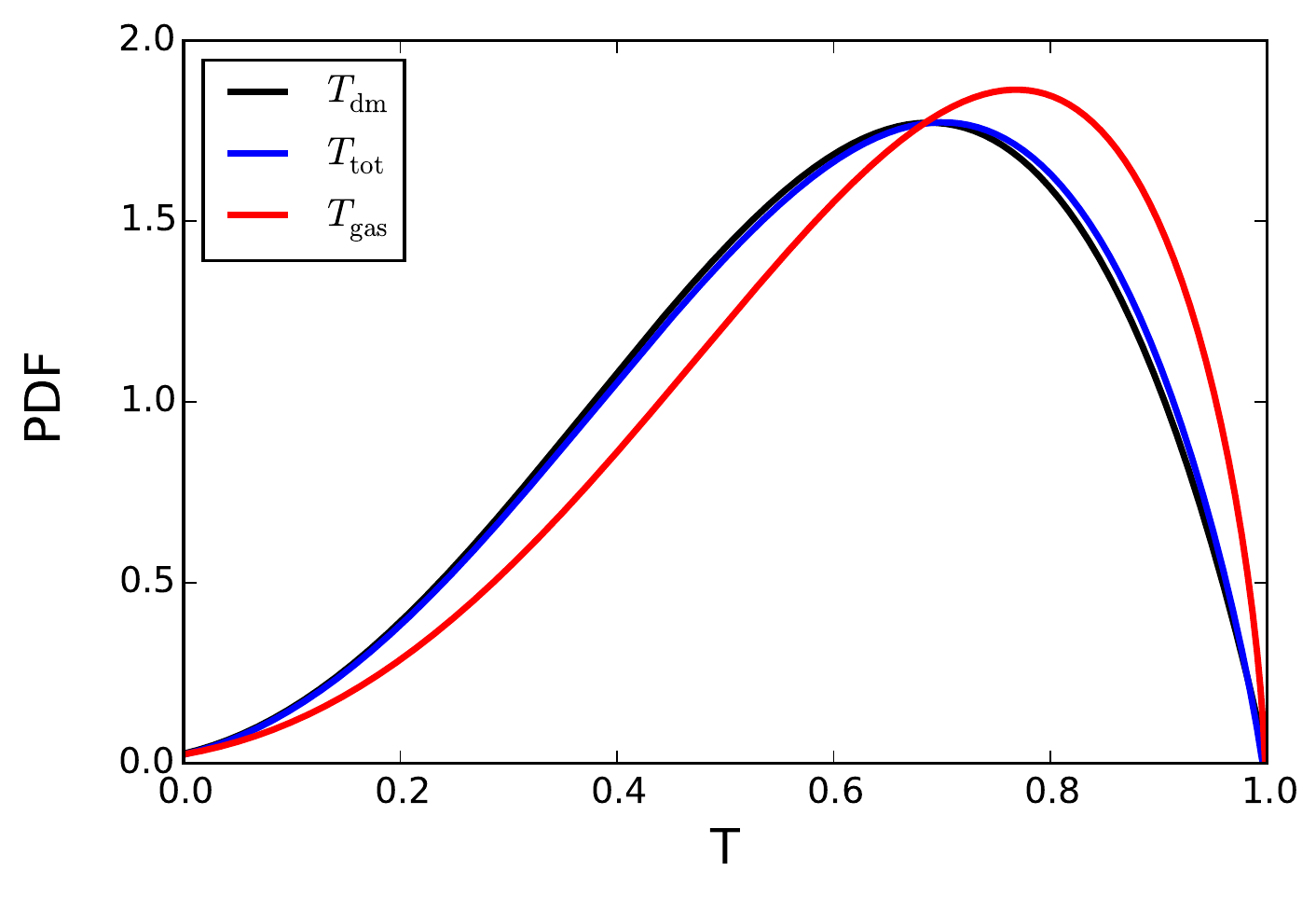}
\caption{Distribution of the triaxiality for the gas (red line), dark matter (black line) and for the combined halo (blue line) at a redshift of $z=14$. The gas is significantly more prolate than the dark matter or the halo as a whole.}  
\label{fig:Triaxparameter of the gas halo}
\end{figure}

We have also investigated the triaxiality of the gas and dark matter components. The best-fit beta-functions to the histograms for gas (red), dark matter (black) and the combined components (blue) are shown in Figure \ref{fig:Triaxparameter of the gas halo}. Once more, we observe a very similar behaviour of the dark matter component compared to the total halo, while the gas deviates slightly. 
This is again due to the fact that the halo is dominated by dark matter. 
For the halo as a whole, we find that the triaxiality distribution peaks at $T_\mathrm{tot} = 0.700 \pm 0.004$ (with $a= 3.35$, $b= 1.89$)  and for the dark matter we recover a very similar value, $T_\mathrm{dm} = 0.688 \pm 0.004$ (with $a= 3.46$, $b= 1.99$). For the gas, we find a shift towards more prolate states with $T_\mathrm{gas} = 0.768 \pm 0.004$ (with $a= 3.69$, $b= 1.69$).

\subsection{Properties of the dense gas component}
\label{Cold gas}
Our simulation not only contains gas and dark matter, but also treats the primordial chemistry accurately. Together with our high resolution throughout the whole simulation volume, this allows us to follow the cooling and gravitational collapse of gas in a large sample of minihaloes, up to densities $n \geq 100 \: {\rm cm^{-3}}$, above which a minihalo could be called star forming. Since the cooling, collapsing gas these minihaloes will eventually form stars, it is worthwhile to examine how closely the spin and shape of the collapsing dense gas component correlate with the properties of the halo as a whole. 

To carry out this analysis, we first identified every minihalo in our $z=14$ snapshot that contained at least 50 gas cells with densities greater than $100 \: {\rm cm^{-3}}$. Imposing this constraint reduced the number of haloes that we consider from 9020 in the original sample to only 
169, but ensured that each halo contained enough dense gas to allow us to draw meaningful conclusions about its angular momentum content. 

We next compute the spin parameter and triaxiality of the gas in each halo with density greater than a threshold value $n_{\rm thres}$ for several different values of the threshold:\footnote{For reference, the mean gas density within a virialized halo at this redshift with the cosmological ratio of baryons to dark matter should be $n_{\rm vir} \sim 0.37 \: {\rm cm^{-3}}$.} $n_{\rm thres} = 1, 10, 50$ and $100 \: {\rm cm^{-3}}$. As previously noted, to do this we first calculate the distance from the centre of the halo to the farthest cell with a density $n \geq n_{\rm thres}$ and then compute the spin parameter and triaxiality for all of the gas within a sphere with a radius equal to this distance. Therefore, in each case our compute values include some contribution from lower density gas, but in practice the dominant contribution to both $\lambda^{\prime}$ and $T$ comes from gas with $n \geq n_{\rm thres}$. 

In a few haloes (especially the most massive ones), we found high density gas at a considerable distance from the centre of the halo, with an intervening low density region separating the two density peaks. These haloes are ones that are caught in the process of merging, and hence appear highly disturbed. In these haloes, we calculate the spin parameter and triaxiality only for the central density peak, as otherwise we recover an artificially high triaxiality and low spin parameter for the dense gas in these merging haloes.

\subsubsection{Spin of the dense gas}
\begin{figure} 
\includegraphics[width=1.10\columnwidth]{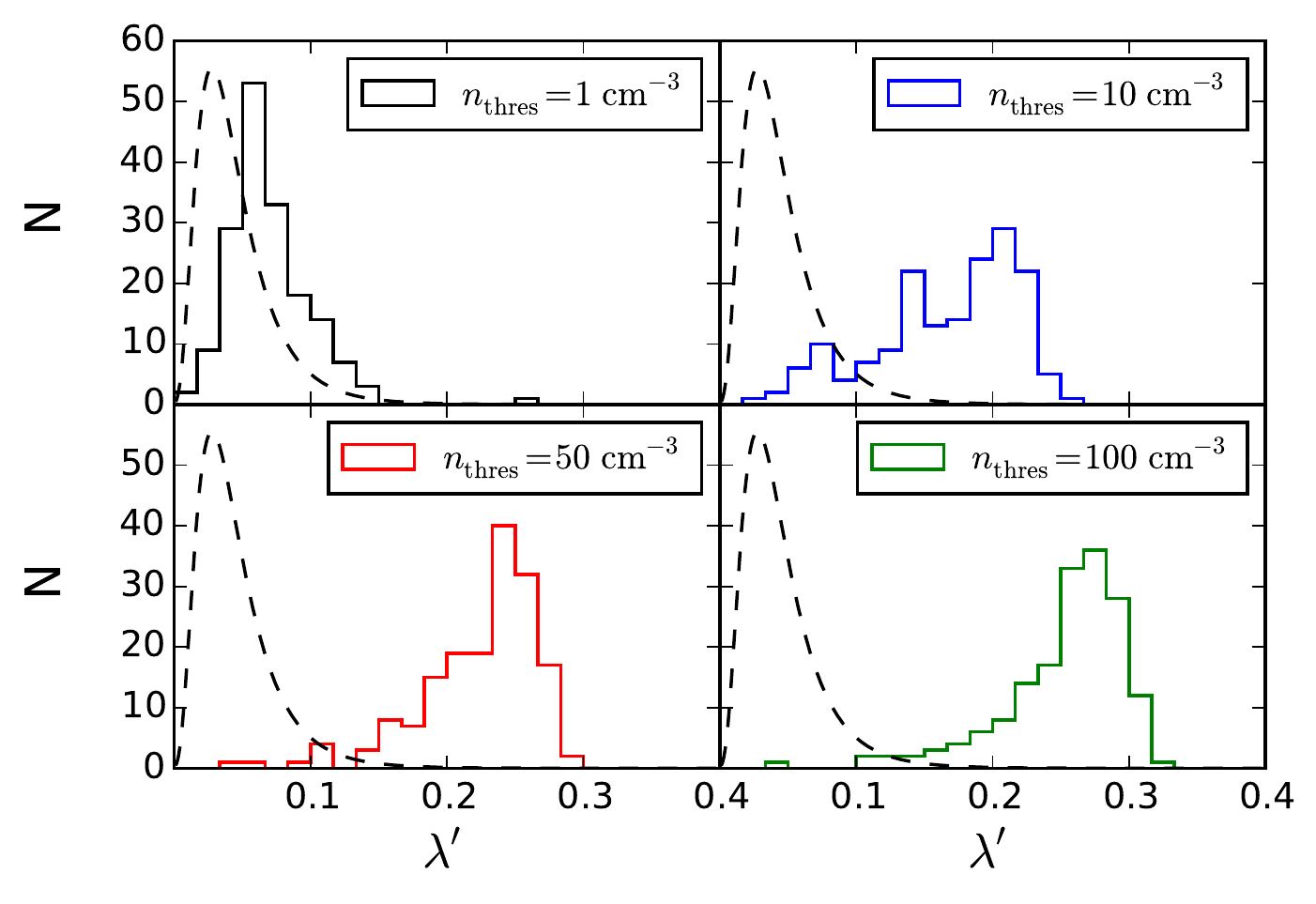}
\caption{Distribution of $\lambda^{\prime}$ in gas denser than $n_{\rm thres}$, shown for $n_{\rm thres} = 1,\,\,10,\,\,50$, and $100$\,cm$^{-3}$. As we increase $n_{\rm thres}$, the peak in the distribution shifts to much larger values of $\lambda^{\prime}$.
The dashed line shows the shape of the distribution for the full sample of 9020 haloes, rescaled by a factor of 169/9020 so that it can be directly compared to the distribution for the sub-sample of haloes containing dense gas.}\label{fig:Lambdadistributioncoldgas}
\end{figure}

In Figure \ref{fig:Lambdadistributioncoldgas}, we show the distributions of $\lambda^{\prime}$ that we obtain for $n_{\rm thres} = 1, 10, 50$ and $100 \: {\rm cm^{-3}}$, which we denote as $\lambda^{\prime}_\mathrm{1}$, $\lambda^{\prime}_\mathrm{10}$, $\lambda^{\prime}_\mathrm{50}$ and $\lambda^{\prime}_\mathrm{100}$, respectively. We see that as we increase $n_{\rm thres}$, the distribution shifts towards substantially larger values of $\lambda^{\prime}$, with the most probable value shifting from 0.025 for our full halo sample to 0.275 for the gas with $n_{\rm thres} = 100 \: {\rm cm^{-3}}$. In addition, the shape of the spin parameter distribution changes -- when $n_{\rm thres}$ is large, it is no longer well-fit by a log-normal function. 
In our relatively small simulation volume, we do not find a single minihalo with very low spin parameter that could be linked to supermassive black hole formation \citep{el95}. However, the number density of these objects is predicted by \citet{el95} to be small (fewer than 1\,Mpc$^{-3}$), and so we would not expect to find any within our simulation volume.

Since the dense gas in our haloes is generally contained within a region with a size that is significantly smaller than the virial radius of the halo, it at first seems plausible that the changes we observe in the distribution of $\lambda^{\prime}$ with increasing $n_{\rm thres}$ may simply reflect the fact that we are studying different spatial scales within our sample of haloes. However, further analysis shows that this does not explain the observed behaviour. Consider, for example, the case of gas with $n \geq 50 \: {\rm cm^{-3}}$. In this case, the typical distance between the centre of the halo and the furthest gas cell is approximately 10\% of the virial radius. If we calculate the spin parameter for each the haloes in our full sample -- even those that contain no dense gas -- at this radius, using the total mass, then we recover a distribution that is very similar to the one we already computed for $R = R_{\rm vir}$, and which differs substantially from the distribution we recover for $n_\mathrm{thres}=50\ \mathrm{cm^{-3}}$ (see Figure \ref{fig:Lambdadistributioncoldgas}). We can therefore conclude that it is the dissipative collapse of the gas that leads to the change in the distribution of $\lambda^{\prime}$, and not simply the fact that we are examining smaller scales.

Furthermore, we examine the density profile of the haloes and obtain on average a power-law distribution with of $\rho \sim r^{-2}$. Similar to \cite{Gao2007} (who has a much better mass resolution), we find no correlation between the slope of the density profile and the spin of a halo.

\subsubsection{Shape of the dense gas}
\begin{figure} 
\includegraphics[width=1.10\columnwidth]{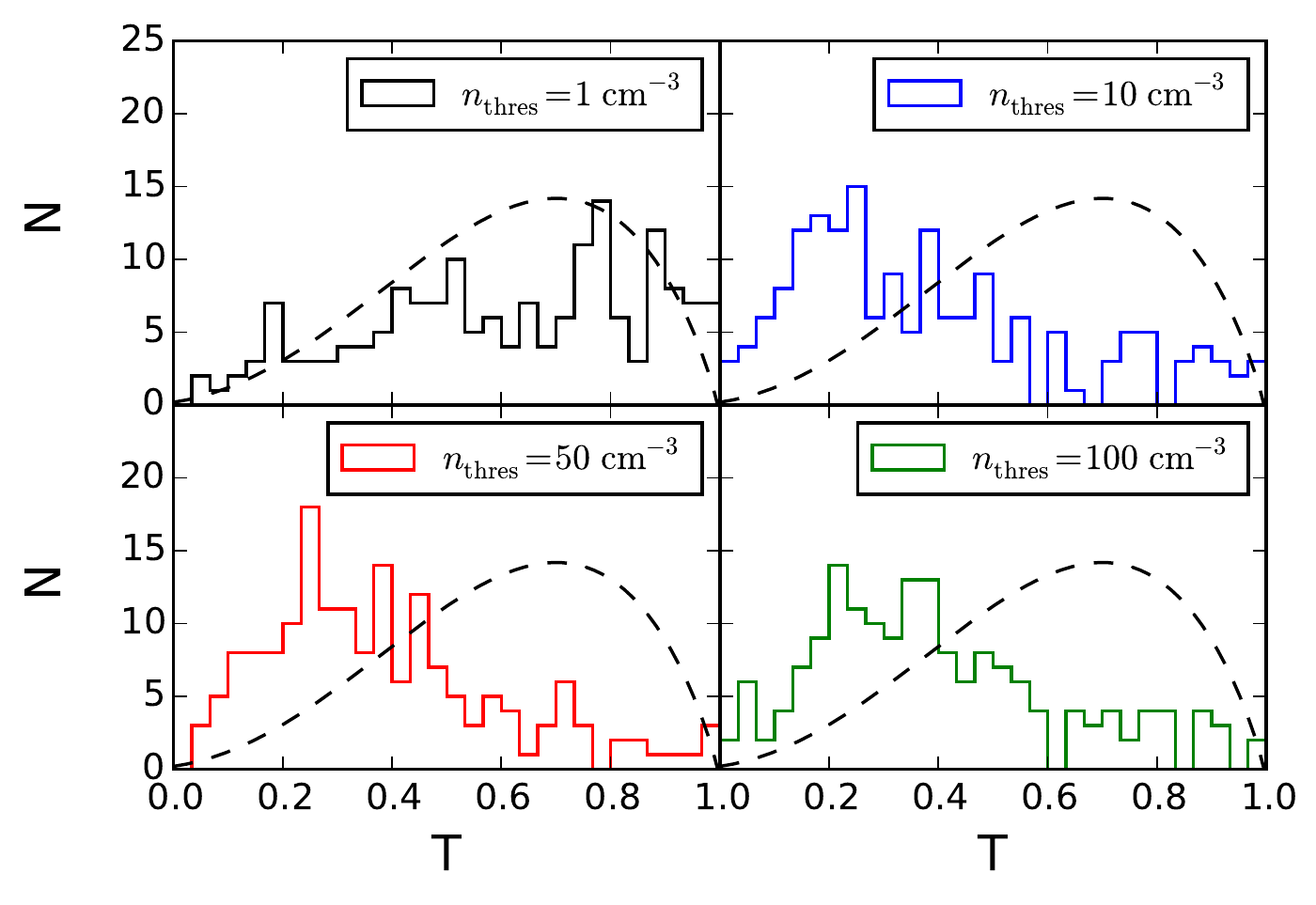}
\caption{As Figure~\ref{fig:Lambdadistributioncoldgas}, but for the triaxiality $T$. Again, the dashed line shows the distribution for the full sample of minihaloes.}
\label{fig:Triaxdistributioncoldgas}
\end{figure}

We also want to investigate whether the shape of the cold, dense gas in the centre deviates from the shape of the halo at the virial radius. To do so, we calculate the triaxiality for the different number density thresholds and plot the resulting distributions in Figure \ref{fig:Triaxdistributioncoldgas}. For comparison, we also show the shape of the triaxiality distribution for our full sample of haloes. 

When $n_\mathrm{thres}=1$\,cm$^{-3}{}$, we find that the gas has a prolate shape in most of the haloes, and that the triaxiality distribution is broadly similar to the distribution for the full halo sample. However, for higher density thresholds, the gas becomes more oblate, as it settles into a rotationally-supported disk-like structure (compare also to Figure \ref{fig:Densityplot}).
This is in agreement with \cite{Gao2007}, who find oblate central gas clouds in most of their halo realizations.

\subsubsection{Correlation between spin and shape}
\begin{figure}
\includegraphics[width=0.99\columnwidth]{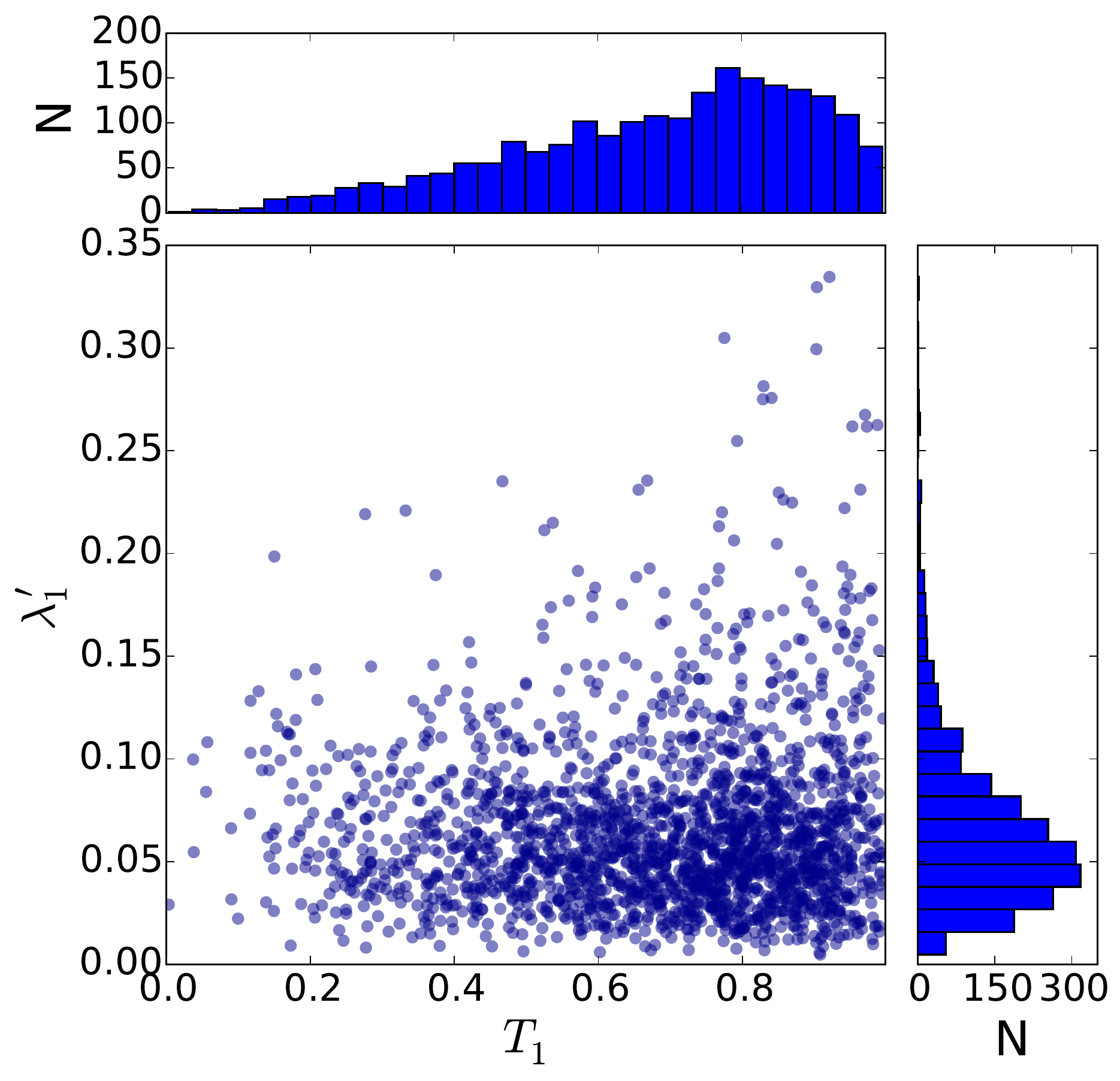}
\caption{Relation between the triaxiality and the spin for gas denser than  $n_\mathrm{thres} = 1 \, {\rm cm^{-3}}$. Histograms of the triaxiality and spin of the gas are shown above and to the right, respectively.}
\label{fig:Triaxiality vs Lambda n=1}
\end{figure}
\begin{figure}
\includegraphics[width=0.99\columnwidth]{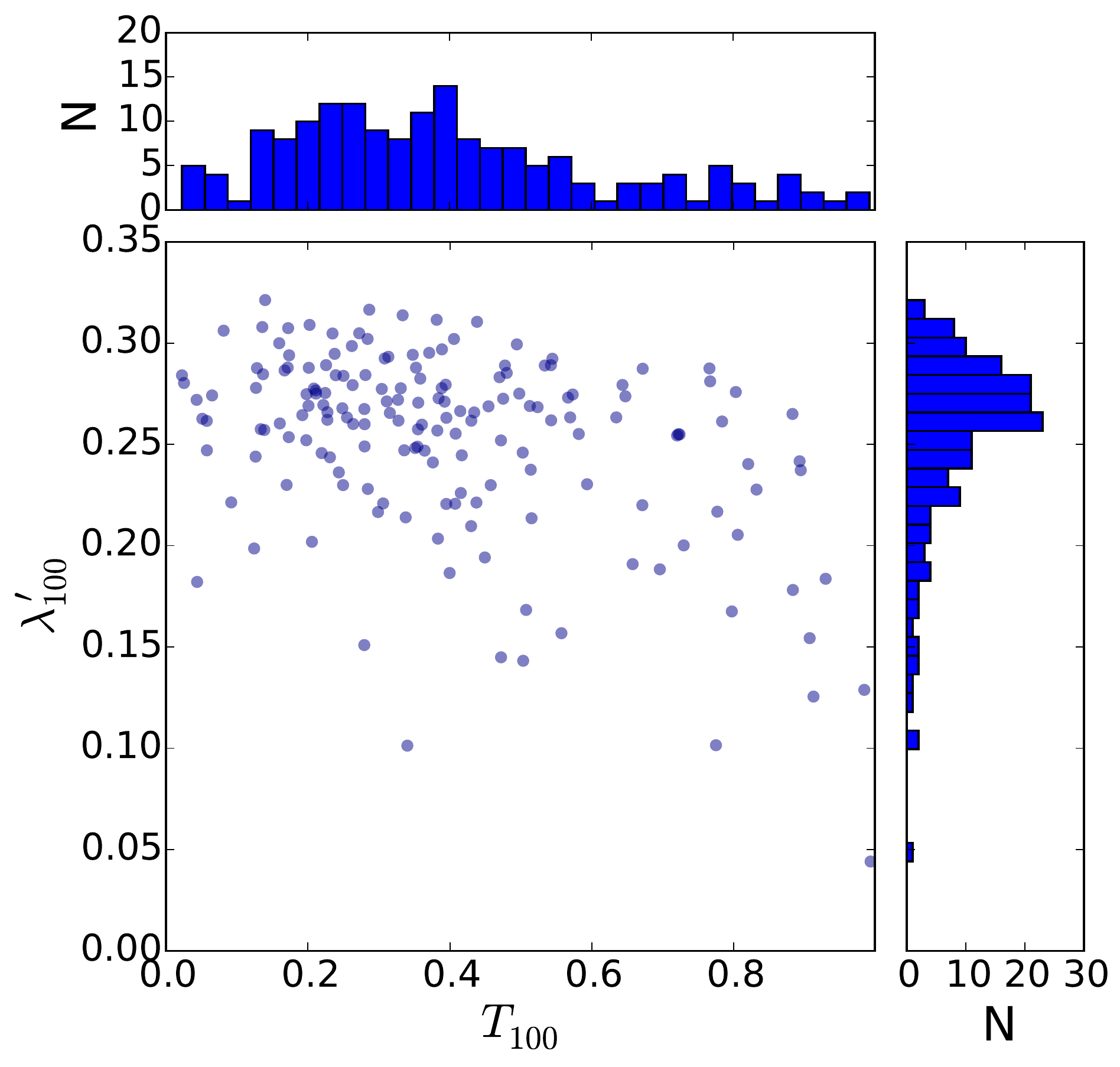}
\caption{As Figure~\ref{fig:Triaxiality vs Lambda n=1}, but for 
$n_\mathrm{thres}=100\ \mathrm{cm}^{-3}$.}  \label{fig:Triaxiality vs Lambda for cold gas with n=100}
\end{figure}
To better understand how the shape and angular momentum of a halo depend on each other, we investigate the correlation of the triaxiality and spin parameter for different number density thresholds. 
In Figures \ref{fig:Triaxiality vs Lambda n=1} and \ref{fig:Triaxiality vs Lambda for cold gas with n=100}, we show the correlation of the spin parameter and triaxiality for cold gas for a low ($n_\mathrm{thres}=1 \: {\rm cm^{-3}}$) and a high ($n_\mathrm{thres}=100 \: {\rm cm^{-3}}$) number density threshold. Both plots are dominated by scatter. For the low number density threshold, we cannot see any trend. 
For the high density threshold, however, a larger triaxiality seems to correlate with a low spin parameter. The interpretation of this result is intuitive: elongated gas clouds have a lower angular momentum than gas clouds that have settled into a (rotating) disk. 

\subsubsection{Does the spin of the dense gas depend on the spin of the halo?}
\begin{figure}
\includegraphics[width=0.99\columnwidth]{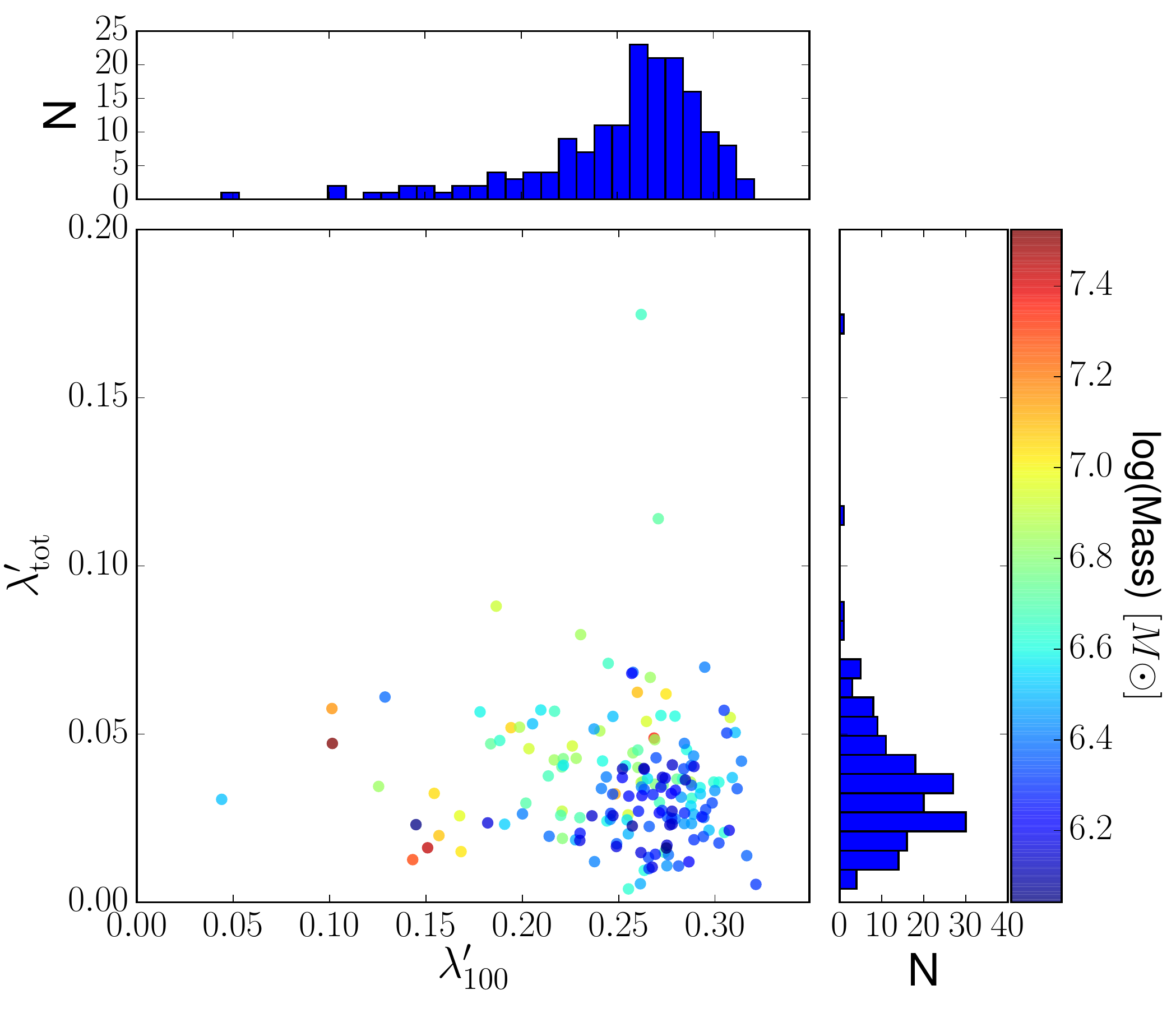}
\caption{Spin parameter of the entire halo plotted against the dense gas spin parameter with a number density threshold $n_{\rm thres} = 100 \, {\rm cm^{-3}}$. The points are colour-coded by the halo mass.} 
\label{fig:Lambd}
\end{figure}

We have already seen that the spin distribution of dense gas in our sample of minihaloes differs significantly from the spin distribution of the full minihalo sample. However, this result says nothing about whether there is a correlation within individual haloes between the spin parameter measured at the virial radius and the spin parameter in the dense, collapsed gas. To explore whether such a correlation exists, we show in Figure~\ref{fig:Lambd} the spin parameter for the halo as a whole ($\lambda^{\prime}_{\rm tot}$) versus the spin parameter for gas with $n > 100 \: {\rm cm^{-3}}$ ($\lambda^{\prime}_{100}$) for each of the haloes in our sample that contain dense gas. The individual points are colour-coded by the halo mass.

The Figure shows quite plainly that there is no correlation between the spin parameter measured on the scale of the virial radius and the spin parameter measured for the dense gas component. Moreover, this result is independent of the halo mass (at least for the range of masses studied here). In addition, we have also investigated whether there is a correlation between the spin of the gas component measured for the whole halo and the spin for $n_{\rm thres} = 100 \: {\rm cm^{-3}}$. We find in this case very similar results to those shown in Figure~\ref{fig:Lambd}, namely that there is no clear correlation between the spin of the gas on the halo scale and the spin of the dense gas component.

One possible concern is that by imposing a density threshold, we may be picking out structures with different scales in different haloes. If $\lambda^{\prime}$ varies monotonically as a function of scale within a given halo, as we would expect if the gas were to conserve angular momentum during its collapse, then by picking out different scales in different haloes, we run the risk of obscuring any underlying correlation. To check that this is not the case, we analysed all of our haloes with dense gas at a scale $R = 0.068\, R_{\rm vir}$, corresponding to the most probable value of $r_{100}$. We find that in this case there is still no correlation between the spin parameter measured on the scale of the virial radius and the spin parameter measured on small scales, meaning that there is no simple way to use the former to predict the latter.

\subsubsection{Spin alignment of the dense gas}
\begin{figure}
\includegraphics[width=0.99\columnwidth]{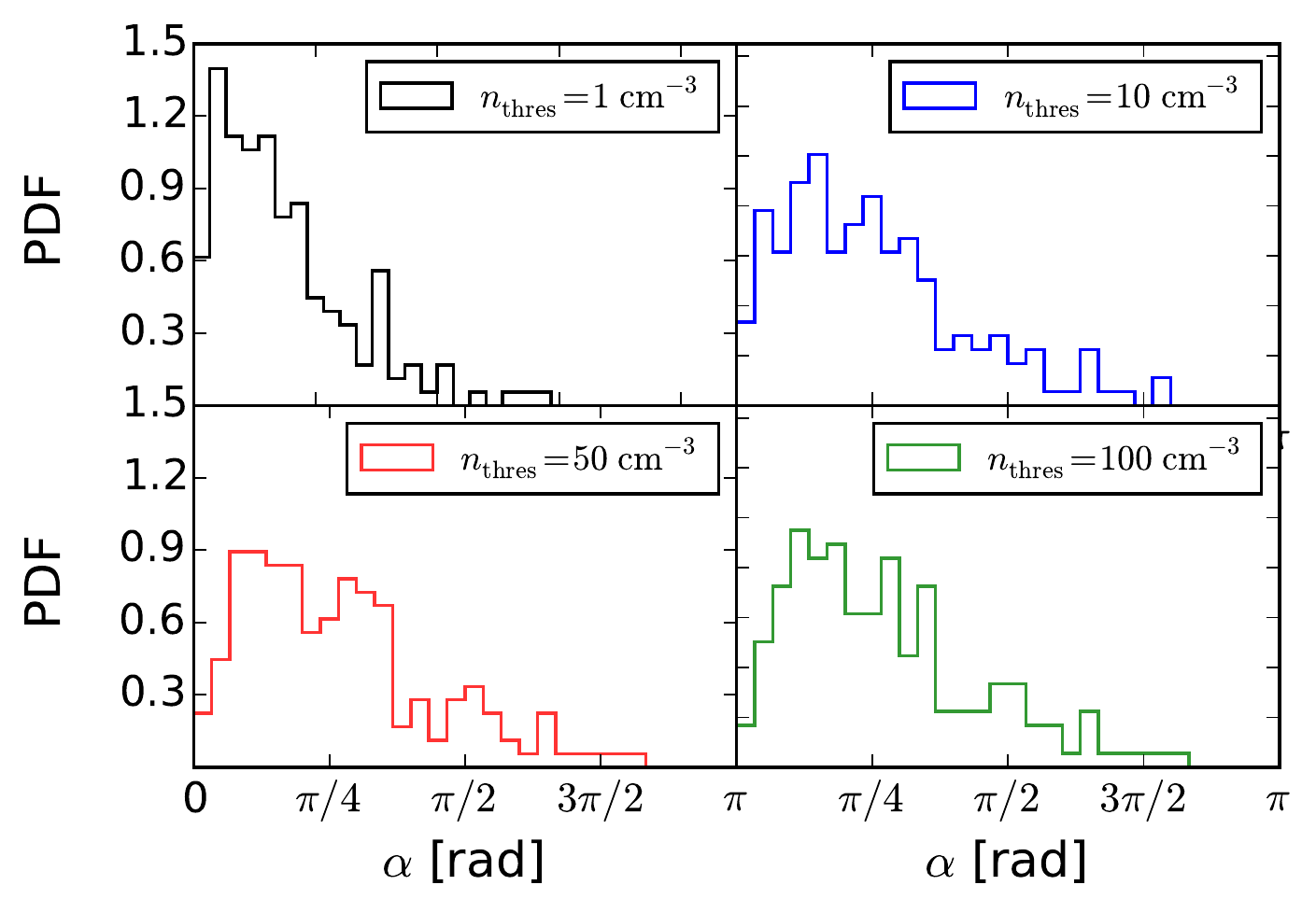}
\caption{Angle between the total gas and dense gas components for several density thresholds.}  \label{fig:Angledistributioncoldgas}
\end{figure}
Finally, we examine whether the angular momentum vector of the dense gas is aligned with that for the total gas. In Figure \ref{fig:Angledistributioncoldgas}, we show the angle between these two vectors for several different values of $n_{\rm thres}$. We find that the rotational axis of the dense component is well aligned with that for the total gas. Although the alignment is not perfect, in most of the haloes containing dense gas the difference in alignments is only around 20$^\circ$. In addition, we see that the distribution of alignments is largely independent of $n_{\rm thres}$, suggesting that the direction of the angular momentum vector does not vary strongly during the collapse of the gas.

\section{Conclusions}\label{Conclusion}
In this study, we have analysed 9020 minihaloes with masses greater than 
$M_{\mathrm{min}} = 6.5 \times 10^{4}\,\mathrm{M}_\odot$, taken from a high resolution cosmological simulation. We have studied the spin and shape of these haloes, as quantified by the spin parameter $\lambda^{\prime}$ and the triaxiality $T$, at a range of different redshifts between $z=24$ and $z=14$. In addition, we have examined how the spin and shape change as we move from considering the halo as a whole to the dense, cooling gas at the centre of the halo that will ultimately form stars. This allows us to investigate whether the properties of these dense sites of future star formation are correlated with the large-scale properties of the halo in which they are located. 

Our main results can be summarized as follows: 
\begin{itemize}
\item The distribution of the spin parameter over all haloes with $M > M_{\rm min}$ can be described by a log-normal distribution with a peak value of $ \mathrm{\lambda^{\prime}} $= 0.0262 $ \pm $ 0.0002 at $z = 14$. 
\item The distribution of triaxialities in this same set of haloes is well described by a beta distribution with parameters $a = 3.35, b = 1.89$ and a peak (i.e.\ most-probable) value of $T = 0.700$ at $z = 14$. The majority of the minihaloes are therefore prolate. 
\item The spin parameter distribution does not evolve significantly with redshift between $z = 24$ and $z = 14$. On the other hand, the triaxiality distribution evolves with redshift, with minihaloes formed at high redshifts being more prolate, on average, than those formed at low redshift.
\item On the scale of the virial radius, the spin and shape of the dark matter component agrees well with the values for the halo as a whole. This is not surprising, since the mass of the halo is dominated by the dark matter. The spin of the gas component on this scale correlates reasonably well with that of the halo, albeit with some scatter, but the gas is typically more prolate than the dark matter.
\item  When we look at dense gas, we see that the spin increases rapidly and becomes more oblate for higher densities. Even for a relatively low density threshold of $ \mathrm{n_{thres}} = 1 \, \mathrm{cm^{-3}}$, the spin distribution cannot be described by a log-normal function. For a much higher number density threshold of $100 \: {\rm cm^{-3}}$, the spin parameter peaks around $\lambda^\prime_{100} \sim 0.25$. 
\item The triaxiality of the gas is largely independent of the spin parameter. However, if we look at gas with a number density $n \geq \mathrm{n_{thres}} = 100 \, \mathrm{cm^{-3}}$, there is a weak inverse correlation between $\lambda^{\prime}$ and the triaxiality: dense gas distributions with smaller spins are more prolate than those that are spinning more rapidly. 
\item The alignments of the spins of nearby haloes are uncorrelated: for halo pairs with separations $\leq 50$~ckpc/$h$, the results are consistent with a random distribution. 
\item There is a good correlation between the alignment of the total angular momentum vector of a halo and the angular momentum of the cold dense gas that it contains, although this correlation weakens as we move to higher densities.
\item The spin parameter of the dense gas is uncorrelated with the spin parameter of the halo as a whole. This means that we cannot use halo-scale measurements of the spin to predict its behaviour on smaller scales.
\end{itemize}

Our results demonstrate that the assumption made by \cite{Souza} in their model for the Population III IMF, namely that the spin of the dense gas on small scales is well correlated with the spin parameter of the halo, is not correct: there is no clear correlation between these quantities. In the future, it would be interesting to follow a statistical ensemble of minihaloes to even higher densities to establish whether we ever reach a point at which the spin parameter distribution becomes independent of density. 

There are several additional directions in which this work could be pursued in the future. For example, the spin and shape of minihaloes has not yet been studied in cosmological simulations of regions of the Universe in which there is a non-zero initial streaming velocity of the baryons with respect to the dark matter, or where there is a non-negligible Lyman-Werner background. It would be interesting to see whether either of these effects has a substantial impact on the spin distribution of the gas or its shape.
In addition, the presence of a dynamically significant magnetic field may also increase the spin parameter of the dense gas \citep{Hirano18}, and so it would be interesting but challenging to carry out a similar study with a magnetohydrodynamical treatment that accounts for the growth of magnetic fields via the turbulent dynamo in these high-redshift minihalos. However, this is out of the scope of our current study.

Another interesting avenue for further study is the relation of the angular momentum of individual minihaloes to the large-scale structure of the Universe or the gas infall rate \citep{sb14}. There are indications from dark matter only simulations that the spin parameter distribution depends on the local clustering of galaxies \citep{Davis2009}. In more clustered regions, mergers will occur more frequently, leading to minihaloes developing larger spins. However, the impact of this on the angular momentum of the gas component is harder to predict and has not been explored on these scales. 

In addition, we know that the spin of present day galaxies is correlated with the underlying cosmic shear field. Tidal torque theory shows that small galaxies tend to align along the filaments in which they form, while larger galaxies show azimuthal orientation \citep[see e.g.][]{codis15, pahwa16}. It would be interesting to see whether this ``spin flip'' can be seen already in minihaloes, which are still in the process of forming from the filaments in the dark matter distribution. We will investigate all of these issues in future work.

\section*{Acknowledgments}
The authors would like to thank the anonymous referee for their helpful and constructive comments.
The authors would like to thank Mattis Magg, Thomas Greif and Daniel Ceverino 
for fruitful discussions. 
The authors acknowledge support from the European Research Council under the European Community's Seventh Framework Programme (FP7/2007 - 2013) via the ERC Advanced Grant ``STARLIGHT: Formation of the First Stars" (project number 339177). SCOG and RSK also appreciate support from the Deutsche Forschungsgemeinschaft via SFB 881, ``The Milky Way System'' (sub-projects
B1, B2 and B8) and SPP 1573 , ``Physics of the Interstellar Medium'' (grant number GL 668/2-1). The authors gratefully acknowledge the Gauss Centre for Supercomputing e.V. (\url{www.gauss-centre.eu}) for providing computing time on the GCS Supercomputer SuperMUC at Leibniz Supercomputing Centre. 
(\url{http://www.lrz.de}). 
The authors acknowledge support by
the state of Baden-W\"urttemberg through bwHPC and the German
Research Foundation (DFG) through grant INST 35/1134-1 FUGG.

\setlength{\bibhang}{2.0em}
\setlength\labelwidth{0.0em}
\bibliographystyle{mn2e}
\bibliography{refs}

\begin{thebibliography}{47}
\expandafter\ifx\csname natexlab\endcsname\relax\def\natexlab#1{#1}\fi

\bibitem[{{Allgood} {et~al}\mbox{.}(2006){Allgood}, {Flores}, {Primack},
  {Kravtsov}, {Wechsler}, {Faltenbacher}, \& {Bullock}}]{Allgood}
{Allgood} B., {Flores} R.~A., {Primack} J.~R., {Kravtsov} A.~V., {Wechsler}
  R.~H., {Faltenbacher} A., {Bullock} J.~S., 2006, \mnras, 367, 1781

\bibitem[{{Barnes} \& {Hut}(1986)}]{bh86}
{Barnes} J., {Hut} P., 1986, \nat, 324, 446

\bibitem[{{Bromm}(2013)}]{volkerreview13}
{Bromm} V., 2013, Reports on Progress in Physics, 76, 112901

\bibitem[{{Bullock} {et~al}\mbox{.}(2001){Bullock}, {Dekel}, {Kolatt},
  {Kravtsov}, {Klypin}, {Porciani}, \& {Primack}}]{Bullock}
{Bullock} J.~S., {Dekel} A., {Kolatt} T.~S., {Kravtsov} A.~V., {Klypin} A.~A.,
  {Porciani} C., {Primack} J.~R., 2001, \apj, 555, 240

\bibitem[{{Clark} {et~al}\mbox{.}(2011){Clark}, {Glover}, {Klessen}, \&
  {Bromm}}]{cgkb11}
{Clark} P.~C., {Glover} S.~C.~O., {Klessen} R.~S., {Bromm} V., 2011, \apj, 727,
  110

\bibitem[{{Clark} {et~al}\mbox{.}(2011b){Clark}, {Glover}, {Smith}, {Greif},
  {Klessen}, \& {Bromm}}]{clark11}
{Clark} P.~C., {Glover} S.~C.~O., {Smith} R.~J., {Greif} T.~H., {Klessen}
  R.~S., {Bromm} V., 2011b, Science, 331, 1040

\bibitem[{{Codis}, {Pichon} \& {Pogosyan}(2015){Codis}, {Pichon}, \&
  {Pogosyan}}]{codis15}
{Codis} S., {Pichon} C., {Pogosyan} D., 2015, \mnras, 452, 3369

\bibitem[{{Davis} \& {Natarajan}(2009)}]{Davis2009}
{Davis} A.~J., {Natarajan} P., 2009, \mnras, 393, 1498

\bibitem[{{Davis} \& {Natarajan}(2010)}]{Davis2010}
{Davis} A.~J., {Natarajan} P., 2010, \mnras, 407, 691

\bibitem[{{Davis} {et~al}\mbox{.}(1985){Davis}, {Efstathiou}, {Frenk}, \&
  {White}}]{fof}
{Davis} M., {Efstathiou} G., {Frenk} C.~S., {White} S.~D.~M., 1985, \apj, 292,
  371

\bibitem[{{de Souza} {et~al}\mbox{.}(2013){de Souza}, {Ciardi}, {Maio}, \&
  {Ferrara}}]{Souza}
{de Souza} R.~S., {Ciardi} B., {Maio} U., {Ferrara} A., 2013, \mnras, 428, 2109

\bibitem[{{Eisenstein} \& {Hu}(1998)}]{eh98}
{Eisenstein} D.~J., {Hu} W., 1998, \apj, 496, 605

\bibitem[{{Eisenstein} \& {Loeb}(1995)}]{el95}
{Eisenstein} D.~J., {Loeb} A., 1995, \apj, 443, 11

\bibitem[{{Franx}, {Illingworth} \& {de Zeeuw}(1991){Franx}, {Illingworth}, \&
  {de Zeeuw}}]{Franx}
{Franx} M., {Illingworth} G., {de Zeeuw} T., 1991, \apj, 383, 112

\bibitem[{{Gao} {et~al}\mbox{.}(2007){Gao}, {Yoshida}, {Abel}, {Frenk},
  {Jenkins}, \& {Springel}}]{Gao2007}
{Gao} L., {Yoshida} N., {Abel} T., {Frenk} C.~S., {Jenkins} A., {Springel} V.,
  2007, \mnras, 378, 449

\bibitem[{{Glover}(2013)}]{glov13}
{Glover} S., 2013, in Astrophysics and Space Science Library, Vol. 396,
  Astrophysics and Space Science Library, {Wiklind} T., {Mobasher} B., {Bromm}
  V., eds., p. 103

\bibitem[{{Glover}(2015)}]{glover15}
{Glover} S.~C.~O., 2015, \mnras, 451, 2082

\bibitem[{{Glover} \& {Abel}(2008)}]{ga08}
{Glover} S.~C.~O., {Abel} T., 2008, \mnras, 388, 1627

\bibitem[{{Glover} \& {Jappsen}(2007)}]{gj07}
{Glover} S.~C.~O., {Jappsen} A.-K., 2007, \apj, 666, 1

\bibitem[{{Greif} {et~al}\mbox{.}(2011){Greif}, {Springel}, {White}, {Glover},
  {Clark}, {Smith}, {Klessen}, \& {Bromm}}]{get11}
{Greif} T.~H., {Springel} V., {White} S.~D.~M., {Glover} S.~C.~O., {Clark}
  P.~C., {Smith} R.~J., {Klessen} R.~S., {Bromm} V., 2011, \apj, 737, 75

\bibitem[{{Hahn} \& {Abel}(2011)}]{hahn11}
{Hahn} O., {Abel} T., 2011, \mnras, 415, 2101

\bibitem[{{Hartwig} {et~al}\mbox{.}(2015a){Hartwig}, {Glover}, {Klessen},
  {Latif}, \& {Volonteri}}]{hartwig15a}
{Hartwig} T., {Glover} S.~C.~O., {Klessen} R.~S., {Latif} M.~A., {Volonteri}
  M., 2015a, \mnras, 452, 1233

\bibitem[{{Hirano} \& {Bromm}(2018)}]{Hirano18}
{Hirano} S., {Bromm} V., 2018, \mnras, 476, 3964

\bibitem[{{Hirano} {et~al}\mbox{.}(2014){Hirano}, {Hosokawa}, {Yoshida},
  {Umeda}, {Omukai}, {Chiaki}, \& {Yorke}}]{Hirano}
{Hirano} S., {Hosokawa} T., {Yoshida} N., {Umeda} H., {Omukai} K., {Chiaki} G.,
  {Yorke} H.~W., 2014, \apj, 781, 60

\bibitem[{{Jang-Condell} \& {Hernquist}(2001)}]{Jang}
{Jang-Condell} H., {Hernquist} L., 2001, \apj, 548, 68

\bibitem[{{Kazantzidis} {et~al}\mbox{.}(2004){Kazantzidis}, {Kravtsov},
  {Zentner}, {Allgood}, {Nagai}, \& {Moore}}]{Kazantzidis}
{Kazantzidis} S., {Kravtsov} A.~V., {Zentner} A.~R., {Allgood} B., {Nagai} D.,
  {Moore} B., 2004, \apjl, 611, L73

\bibitem[{{Knebe} {et~al}\mbox{.}(2009){Knebe}, {Wagner}, {Knollmann},
  {Diekershoff}, \& {Krause}}]{Knebe}
{Knebe} A., {Wagner} C., {Knollmann} S., {Diekershoff} T., {Krause} F., 2009,
  \apj, 698, 266

\bibitem[{{McKee} \& {Tan}(2008)}]{McKee}
{McKee} C.~F., {Tan} J.~C., 2008, \apj, 681, 771

\bibitem[{{Mo}, {Mao} \& {White}(1998){Mo}, {Mao}, \& {White}}]{Mo}
{Mo} H.~J., {Mao} S., {White} S.~D.~M., 1998, \mnras, 295, 319

\bibitem[{{Mocz} {et~al}\mbox{.}(2015){Mocz}, {Vogelsberger}, {Pakmor},
  {Genel}, {Springel}, \& {Hernquist}}]{mocz15}
{Mocz} P., {Vogelsberger} M., {Pakmor} R., {Genel} S., {Springel} V.,
  {Hernquist} L., 2015, \mnras, 452, 3853

\bibitem[{{Pahwa} {et~al}\mbox{.}(2016){Pahwa}, {Libeskind}, {Tempel},
  {Hoffman}, {Tully}, {Courtois}, {Gottl{\"o}ber}, {Steinmetz}, \&
  {Sorce}}]{pahwa16}
{Pahwa} I. {et~al.}, 2016, \mnras, 457, 695

\bibitem[{{Pakmor} {et~al}\mbox{.}(2016){Pakmor}, {Springel}, {Bauer}, {Mocz},
  {Munoz}, {Ohlmann}, {Schaal}, \& {Zhu}}]{pakmor16}
{Pakmor} R., {Springel} V., {Bauer} A., {Mocz} P., {Munoz} D.~J., {Ohlmann}
  S.~T., {Schaal} K., {Zhu} C., 2016, \mnras, 455, 1134

\bibitem[{{Peebles}(1969)}]{Peebles}
{Peebles} P.~J.~E., 1969, \apj, 155, 393

\bibitem[{{Planck Collaboration} {et~al}\mbox{.}(2016){Planck Collaboration},
  {Ade}, {Aghanim}, {Arnaud}, {Ashdown}, {Aumont}, {Baccigalupi}, {Banday},
  {Barreiro}, {Bartlett}, \& et~al.}]{planck15}
{Planck Collaboration} {et~al.}, 2016, \aap, 594, A13

\bibitem[{{Sasaki} {et~al}\mbox{.}(2014){Sasaki}, {Clark}, {Springel},
  {Klessen}, \& {Glover}}]{Sasaki}
{Sasaki} M., {Clark} P.~C., {Springel} V., {Klessen} R.~S., {Glover} S.~C.~O.,
  2014, \mnras, 442, 1942

\bibitem[{{Schauer} {et~al}\mbox{.}(2018){Schauer}, {Glover}, {Klessen}, \&
  {Ceverino}}]{anna18}
{Schauer} A.~T.~P., {Glover} S.~C.~O., {Klessen} R.~S., {Ceverino} D., 2018, in
  prep.

\bibitem[{{Schauer} {et~al}\mbox{.}(2017){Schauer}, {Regan}, {Glover}, \&
  {Klessen}}]{anna17b}
{Schauer} A.~T.~P., {Regan} J., {Glover} S.~C.~O., {Klessen} R.~S., 2017,
  \mnras, 471, 4878

\bibitem[{{Smith} {et~al}\mbox{.}(2011){Smith}, {Glover}, {Clark}, {Greif}, \&
  {Klessen}}]{sm11}
{Smith} R.~J., {Glover} S.~C.~O., {Clark} P.~C., {Greif} T., {Klessen} R.~S.,
  2011, \mnras, 414, 3633

\bibitem[{{Springel}(2010)}]{arepo}
{Springel} V., 2010, \mnras, 401, 791

\bibitem[{{Springel} \& {Hernquist}(2002)}]{gadget2}
{Springel} V., {Hernquist} L., 2002, \mnras, 333, 649

\bibitem[{{Springel}, {White} \& {Hernquist}(2004){Springel}, {White}, \&
  {Hernquist}}]{Springel2004}
{Springel} V., {White} S.~D.~M., {Hernquist} L., 2004, in IAU Symposium, Vol.
  220, Dark Matter in Galaxies, {Ryder} S., {Pisano} D., {Walker} M., {Freeman}
  K., eds., p. 421

\bibitem[{{Stacy} \& {Bromm}(2014)}]{sb14}
{Stacy} A., {Bromm} V., 2014, \apj, 785, 73

\bibitem[{{Stacy}, {Greif} \& {Bromm}(2010){Stacy}, {Greif}, \&
  {Bromm}}]{stacy10}
{Stacy} A., {Greif} T.~H., {Bromm} V., 2010, \mnras, 403, 45

\bibitem[{{Stacy}, {Greif} \& {Bromm}(2012){Stacy}, {Greif}, \&
  {Bromm}}]{stacy12}
{Stacy} A., {Greif} T.~H., {Bromm} V., 2012, \mnras, 422, 290

\bibitem[{{Teklu} {et~al}\mbox{.}(2015){Teklu}, {Remus}, {Dolag}, {Beck},
  {Burkert}, {Schmidt}, {Schulze}, \& {Steinborn}}]{Teklu}
{Teklu} A.~F., {Remus} R.-S., {Dolag} K., {Beck} A.~M., {Burkert} A., {Schmidt}
  A.~S., {Schulze} F., {Steinborn} L.~K., 2015, \apj, 812, 29

\bibitem[{{Warren} {et~al}\mbox{.}(1992){Warren}, {Quinn}, {Salmon}, \&
  {Zurek}}]{Warren}
{Warren} M.~S., {Quinn} P.~J., {Salmon} J.~K., {Zurek} W.~H., 1992, \apj, 399,
  405

\bibitem[{{Wise} {et~al}\mbox{.}(2012){Wise}, {Turk}, {Norman}, \&
  {Abel}}]{wise12}
{Wise} J.~H., {Turk} M.~J., {Norman} M.~L., {Abel} T., 2012, \apj, 745, 50

\end{thebibliography}
\label{lastpage}

\end{document}